\documentclass[a4paper,fleqn,usenatbib]{mnras}
\usepackage[T1]{fontenc}
\usepackage{ae,aecompl}
\usepackage{graphicx}	
\usepackage{amsmath}	
\usepackage{amssymb}	
\usepackage{color}
\usepackage{array}
\usepackage{lineno}
\usepackage{bookmark}
\title[Jet Structure in the Afterglow Phase]
{Jet Structure in the Afterglow Phase for Gamma-ray Bursts with a Precessing Jet}
\author[Huang et al.]{
\parbox{\textwidth}{
Bao-Quan Huang,$^{1}$
Da-Bin Lin,$^{1}$\thanks{E-mail: lindabin@gxu.edu.cn}
Tong Liu,$^{2}$
Jia Ren,$^{1}$
Xiang-Gao Wang,$^{1}$\\
Hong-Bang Liu,$^{1}$
En-Wei Liang$^{1}$\\}\\
$^1$Laboratory for Relativistic Astrophysics, Department of Physics, Guangxi University, Nanning 530004, China\\
$^2$Department of Astronomy, Xiamen University, Xiamen, Fujian 361005, China\\}
\date{Accepted XXX. Received YYY; in original form ZZZ}
\pubyear{2015}
\begin{document}
\label{firstpage}
\pagerange{\pageref{firstpage}--\pageref{lastpage}}
\maketitle

\begin{abstract}
The structured jet is involved to explain the afterglows and even the prompt emission of GRB~170817A.
In this paper, we stress that for a precessing jet,
the jet structure in the prompt emission phase and that in the afterglow phase may be different.
The jet structure in the afterglow phase can be non-uniform
even if a narrow-uniform jet is presented in the prompt emission phase.
We estimate the jet structure in the afterglow phase under the situation
that a narrow-uniform-precessing jet is launched from the central engine of gamma-ray burst.
With different precession angles,
it is found that
the structured jet can be roughly described as follows:
a narrow uniform core with power-law wings and sharp cut-off edges,
a Gaussian profile, a ring shape, or other complex profile in energy per solid angle.
Correspondingly, the afterglows for our obtained structured jets are also estimated.
We find that the estimates of the intrinsic kinetic energy,
the electron index, and the jet opening angle
based on the afterglows formed in a precessing system may be incorrect.
Our obtained structured jet is likely to be revealed by future observations for
a fraction of gravitational wave detected merging compact binary systems (e.g., black hole-neutron star mergers).
\end{abstract}
\section{Introduction}\label{Sec:Intro}

Gamma-ray bursts (GRBs) are the most powerful electromagnetic explosions in the universe.
They are widely argued to originate from the compact binaries mergers or core collapse of massive stars.
On August 17, 2017 at 12:41:04 UTC,
the advanced Laser Interferometer Gravitational-wave Observatory and the
Advanced Virgo gravitational-wave detectors made their first detection
of a gravitational wave event (GW~170817) from the merger of a binary neutron star system (\citealp{2017ApJ...850L..39A}; \citealp{2017ApJ...850L..40A};
\citealp{2017ApJ...841...89A}; \citealp{2017PhRvD..96b2001A}).
About 2~s post-merger,
the Fermi (\citealp{2017ApJ...848L..14G}) and INTEGRAL satellites (\citealp{2016ApJ...820L..36S})
observed a short burst (GRB~170817A)
from a location coincident with GW~170817.
A  series of observation campaigns following this discovery
have led to the detection of
a bright optical counterpart,
AT2017gfo (\citealp{2017Natur.551...64A};
\citealp{2017Sci...358.1556C};
\citealp{2017Natur.551...80K};
\citealp{2017Sci...358.1559K};
\citealp{2017Natur.551...67P};
\citealp{2017Natur.551...75S};
\citealp{2017ApJ...848L..16S};
\citealp{2017ApJ...848L..24V};
\citealp{2017SciBu..62.1433H};
\citealp{2017ApJ...848L..27T};
\citealp{2017ApJ...850L...1L})
associated with the kilonova powered by the radioactive decay
of heavy elements formed in the binary neutron star merger
(\citealp{1998ApJ...507L..59L};
\citealp{2012ApJ...746...48M};
\citealp{2013ApJ...774L..23B};
\citealp{2016ARNPS..66...23F};
\citealp{2018MNRAS.477.2173S};
\citealp{2017NewAR..79....1L}).
The joint GW-GRB detection has provided the first compelling observational evidence on the relation of short
GRBs and the binary neutron star mergers.

The physical origin of GRB~170817A emission is still under debate.
The prompt $\gamma$-rays of GRB~170817A is argued to originate from the photosphere of jets (\citealp{2018ApJ...860...72M}),
the internal shocks (\citealp{2017ApJ...848L..34M}; \citealp{2017arXiv171008514F}),
the internal-collision-induced magnetic reconnection and turbulence
(\citealp{2018ApJ...860...72M}; \citealp{2011ApJ...726...90Z}),
or the external-reverse shock (\citealp{2017arXiv171008514F}).
Meanwhile, a diverse sets of jet structure are involved, e.g.,
an off-axis top-hat jet (\citealp{2018ApJ...856...90L}),
a mildly relativistic and isotropic fireball,
and a structured jet (e.g.,
\citealp{2018A&A...613L...1D};
\citealp{2018MNRAS.478..733L};
\citealp{2018PhRvL.120x1103L};
\citealp{Lyman_JD-2018-Lamb_GP-NatAs.2.751L};
\citealp{2018ApJ...856L..18M};
 \citealp{2018ApJ...860...72M};
 \citealp{2018ApJ...867...57R};
\citealp{2018MNRAS.478L..18T}).
These models should explain both the radiation spectrum and
the lag ($t_{\rm lag}\sim 2\rm s$) of GRB prompt emission
relative to GW~170817.
An off-axis top-hat jet launched immediately after the merger
can naturally produce the relation of $t_{\rm lag}\sim T_{90}$
with $T_{90}$ being the duration of prompt $\gamma$-ray emission (\citealp{2018ApJ...856...90L}).
However, $t_{\rm lag}$ is mainly caused by the delay of the merger and jet launching
under the framework of photosphere model (\citealp{Zhang_BB-2018-Zhang_B-NatCo.9.447Z,2018ApJ...860...72M}) or
is corresponding to the shock breakout from a cocoon (\citealp{2018MNRAS.479..588G}).
The prompt emission of GRB~170817A may be the scattered emission of a short GRB by a cocoon
and the values of $t_{\rm lag}$ and $T_{90}$ are reproduced with typical short GRB parameters (\citealp{2018ApJ...867...39K}).
Apart from the prompt emission, the broad-band afterglows of GRB~170817A is peculiar
but such kind of afterglows was predicted before GW~170817/GRB~170817A (\citealp{2017MNRAS.472.4953L}).
Nonetheless, the origin of the outflow structure is still open to debate.
The afterglows were first detected at $\sim$9 days
by Chandra in the X-rays
(\citealp{2017ApJ...848L..25H};
\citealp{2017Sci...358.1565E};
\citealp{2018ApJ...856L..18M};
\citealp{2017Natur.551...71T})
and $\sim$16 days in the radio band
(\citealp{2017Sci...358.1579H})
after GW~170817.
Its spectrum can be described as a single power-law
for radio-optical-X-ray observations (\citealp{Lyman_JD-2018-Lamb_GP-NatAs.2.751L}),
which is consistent with the emission from a relativistic external-forward shock.
However, the afterglows continued to rise in flux until $\gtrsim115$ days after GW~170817
(e.g.,
\citealp{Lyman_JD-2018-Lamb_GP-NatAs.2.751L};
\citealp{2018ApJ...856L..18M};
\citealp{2018Natur.554..207M};
\citealp{2018ApJ...853L...4R};
\citealp{2017Natur.551...71T}).
The continued brightening is anomalous for canonical GRBs
and in favor of the structured jet scenario.
Recently,
\cite{2018MNRAS.478L..18T} found that
a Gaussian profile jet with an off-axis observer
is successful in capturing the observed features
of GRB~170817A for a year-long afterglow monitoring
(see also \citealp{Lamb_GP-2019-Lyman_JD-ApJ.870L.15L,vanEerten_ETH-2018-Ryan_G-arXiv180806617V}).
Thus, the structured jets are involved to explain the afterglows
and even the prompt $\gamma$-rays of GRB~170817A (see the discussion in the beginning of this paragraph).
In this paper,
however,
we would like to point out that the jet structure in the prompt emission phase may be very different
from that in the afterglow emission phase,
especially in GRBs with a precessing jet.
For GRBs with a narrow-uniform-precessing jet,
a structured jet is difficult to form in the prompt emission phase
due to the low frequency of mergers between jet shells.
In the afterglow phase,
the early launched jet shells are decelerated during its propagation into the circum-burst medium.
Thus, the later launched jet shells can catch up and collide with
the early launched ones in the early phase of afterglow.
In the situation with a precessing jet,
a structured jet can be easily formed in the afterglow phase.

We study the jet structure in the afterglow phase under the situation that a precessing jet is launched
from the central engine of GRBs.
The paper is organized as follows.
The procedure to calculate the jet structure and the obtained structured jet are shown
in Sections~\ref{Sec:method} and \ref{Sec:Result}, respectively.
The conclusions and discussion are presented in Section~\ref{Sec:Conclusion_and_Discussion}.
\section{Procedure to Calculate the Jet Structure}\label{Sec:method}
\begin{figure}
\centering
\includegraphics[width=.3\textwidth]{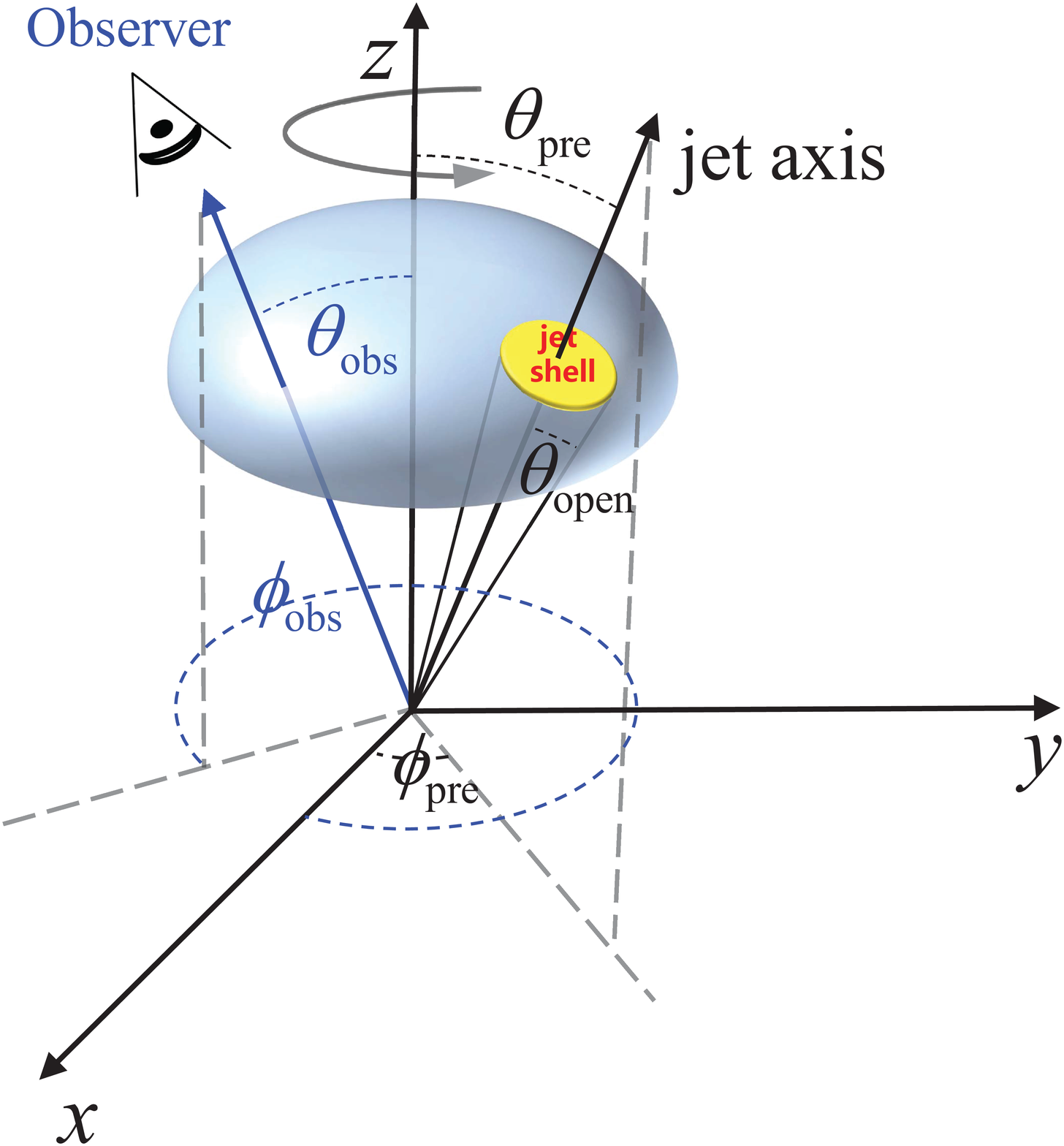}
\caption{Schematic picture of a precessing jet in GRBs.}
\label{myFigA}
\end{figure}

Jet precession has been previously discussed as a phenomenon relevant for GRBs
(e.g., \citealp{1996ApJ...473L..79B}; \citealp{1999ApJ...520..666P}; \citealp{2006A&A...454...11R}; \citealp{2007A&A...468..563L}; \citealp{2011PhRvD..83b4005F}; \citealp{Stone_N-2013-Loeb_A-PhRvD.87h4053S}; \citealp{2017NewAR..79....1L}; \citealp{2018MNRAS.474L..81L}).
The large amplitude precession requires large amplitude
misalignment of the post-merger black hole (BH) and its accretion disk.
In BH-neutron star (NS) mergers, the BH may possess a larger natal reservoir of spin angular momentum,
allowing for greater misalignment between the post-merger BH and  the disk formed from NS debris.
Then, \cite{Stone_N-2013-Loeb_A-PhRvD.87h4053S} stated
that the precession can carry significant observational consequences for BH-NS mergers.
However, the situation for NS-NS mergers is different except a millisecond spin NS is involved in the binaries (\citealp{Stone_N-2013-Loeb_A-PhRvD.87h4053S}).
GW~170817 is the first detected GW signal from a NS-NS merger
(but see \citealp{2018arXiv180803836H} for BH-NS merger origin of GW~170817).
Then, the precession may be likely too small to carry significant observational consequences
in GRB~170817A.
The detail mechanisms being response to the jet precession can refer to \cite{Stone_N-2013-Loeb_A-PhRvD.87h4053S}.
If a spinning black hole is surrounded by a tilted accretion disk,
a precessing jet may be launched from the central engine of GRBs.
The schematic picture is shown in Figure \ref{myFigA},
where the central engine of GRBs locates at the origin of coordinate ($r=0$).
The yellow region represents the narrow-uniform-precessing jet with opening angle $\theta_{\rm open}$,
which rotates around $z$-axis with a precession angle $\theta_{\rm pre}$.
The spherical coordinate $(r, \theta, \phi)$
with $\theta=0$ being along the direction of $z$-axis is used in this paper,
$(\theta_{\rm obs},\phi_{\rm obs})$ is adopted to describe the direction of the observer,
and $(\theta_{\rm pre}, \phi_{\rm pre})$ represents the direction of the jet axis at time $t$.
Due to the jet precession,
the total output energy of jet would be distributed into a large solid angle.
This behavior depends on the precession period $\tau$ and the evolution behavior of the jet power $P(t)$.

In this work, we compute the final jet structure, i.e., the energy distribution $\varepsilon(\theta,\varphi)$,
for a narrow-uniform-precessing jet. The procedures to obtain the jet structure are shown as follows.
We divide the time $t$ into a series of time intervals $[0, \delta t]$, $[\delta t,2\delta t]$, $\ldots$, $[(k-1)\delta t,k\delta t]$, $\ldots$, $[t_{\rm end}-\delta t,t_{\rm end}]$
with $t_{\rm end}=K\delta t$ being the duration of jet activities.
For each time interval,
the total output energy of the jet can be estimated with $P(t=k\delta t-\delta t/2)\delta t$,
which will be redistributed to $n$ infinitesimal-ejecta-cells (IECs).
Here, the solid angle of an IEC is set to zero and the energy of an IEC is $P(t=k\delta t-\delta t/2)\delta t/n$, which is different for different time interval since $P(t)$ is a time-dependent function.
In the $k$th time interval,
we launch $N=2n/(1-\cos \theta _{\rm open})$ IECs
with randomly selected direction $(\theta_{\rm cell}, \phi_{\rm cell})$,
where the value of $\cos \theta_{\rm cell}$ and $\phi_{\rm cell}$ are randomly took
in the range of $[-1,1]$ and $[0^{\circ}, 360^{\circ}]$, respectively.
For these $N(\gg1)$ cells,
only $\sim n$ IECs fall into the solid angle of the jet shell at time $t=(k-1/2)\delta t$
and are used to represent the energy output of the jet in the $k$th time interval.
We select out these $\sim n$ IECs by using the following relation:
\begin{eqnarray}\label{Eq:Condition}
\begin{array}{l}
\cos {\theta _{{\rm{open}}}} \leqslant \sin {\theta _{{\rm{cell}}}}\cos {\phi _{{\rm{cell}}}}\sin {\theta _{{\rm{pre}}}}\cos {\phi _{{\rm{pre}}}} + \\
\;\;\;\;\;\;\;\sin {\theta _{{\rm{cell}}}}\sin {\phi _{{\rm{cell}}}}\sin {\theta _{{\rm{pre}}}}\sin {\phi _{{\rm{pre}}}} + \cos {\theta _{{\rm{cell}}}}\cos {\theta _{{\rm{pre}}}},
\end{array}
\end{eqnarray}
where $(\theta_{\rm pre}, \phi_{\rm pre})$ describes the direction of the jet axis at time $t=(k-1/2)\delta t$.
If $N\gg1$ is satisfied, the total energy of our selected IECs at time $t=(k-1/2)\delta t$
would be pretty close to $P(t)\delta t$, which is the total output energy in the $k$th time interval.
By changing $k$ from 1 to $K$, we can select out a series of IECs with different energy and different direction.
We divide the solid angle of $\theta\in [\max\{\theta_{\rm pre}-\theta_{\rm open},0\}, \min\{\theta_{\rm pre}+\theta_{\rm open},\pi/2\}]$ and $\phi\in [0, 2\pi]$ into $500\times 200$ grids.
Our selected IECs would fall into different grids.
The total energy of a grid is obtained by summing up the energy of these IECs
which fall into the corresponding grid.
By dividing the grid's total energy to the grid's solid angle,
one can obtain the energy density $\varepsilon$ per solid angle for grids.

\section{Results}\label{Sec:Result}
\begin{figure*}
\centering\scriptsize{
\begin{tabular}{|c|c|c|}
\hline
$\;\;\;\theta_{\rm pre}=2.5^\circ$ & $\;\;\;\theta_{\rm pre}=5^\circ$ & $\;\;\;\theta_{\rm pre}=10^\circ$ \\
\includegraphics[scale=0.28,trim=45 265 45 138,clip]{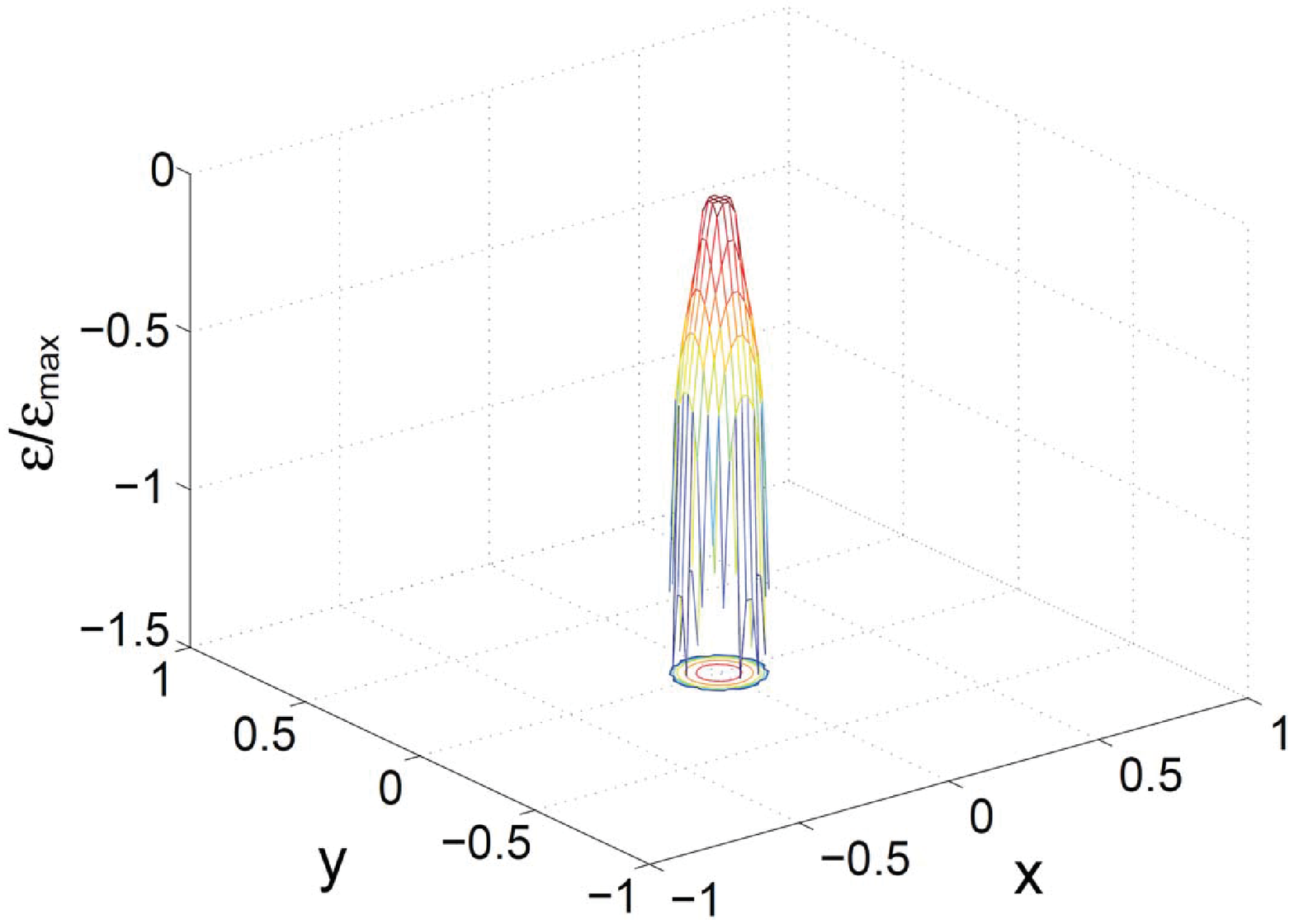} &
\includegraphics[scale=0.28,trim=45 265 45 138,clip]{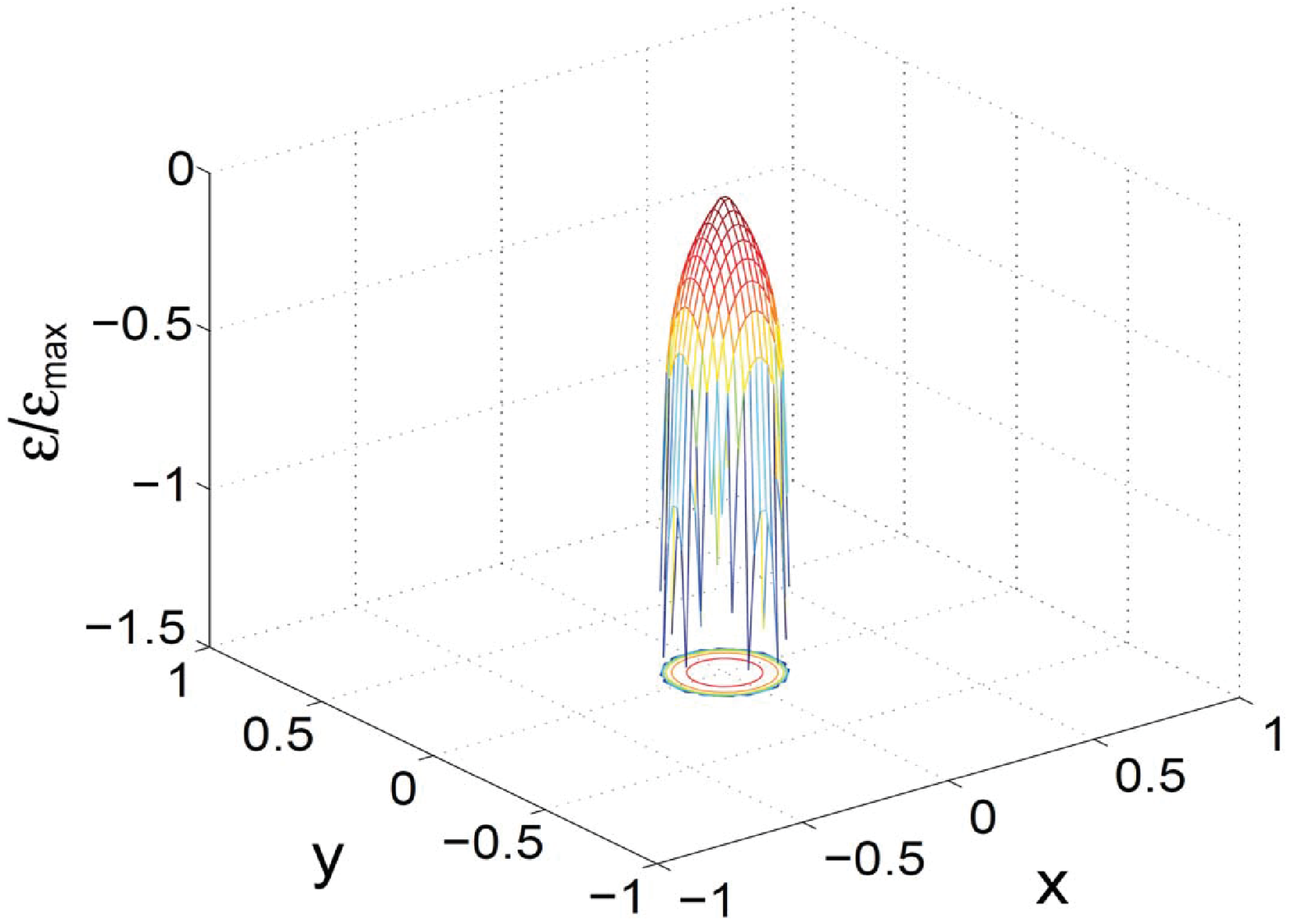} &
\includegraphics[scale=0.28,trim=45 265 45 138,clip]{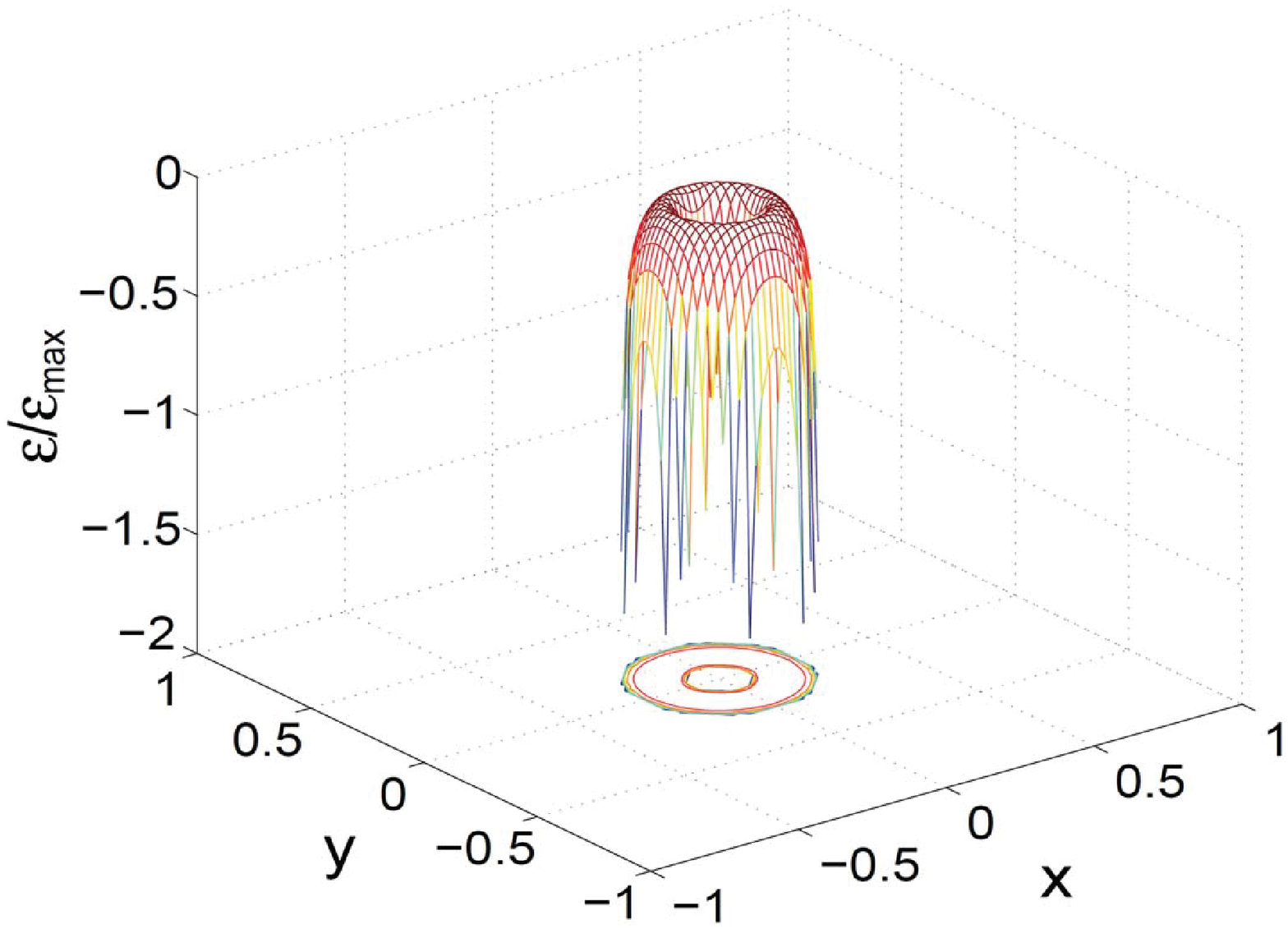} \\
\includegraphics[scale=0.28,trim=45 265 45 138,clip]{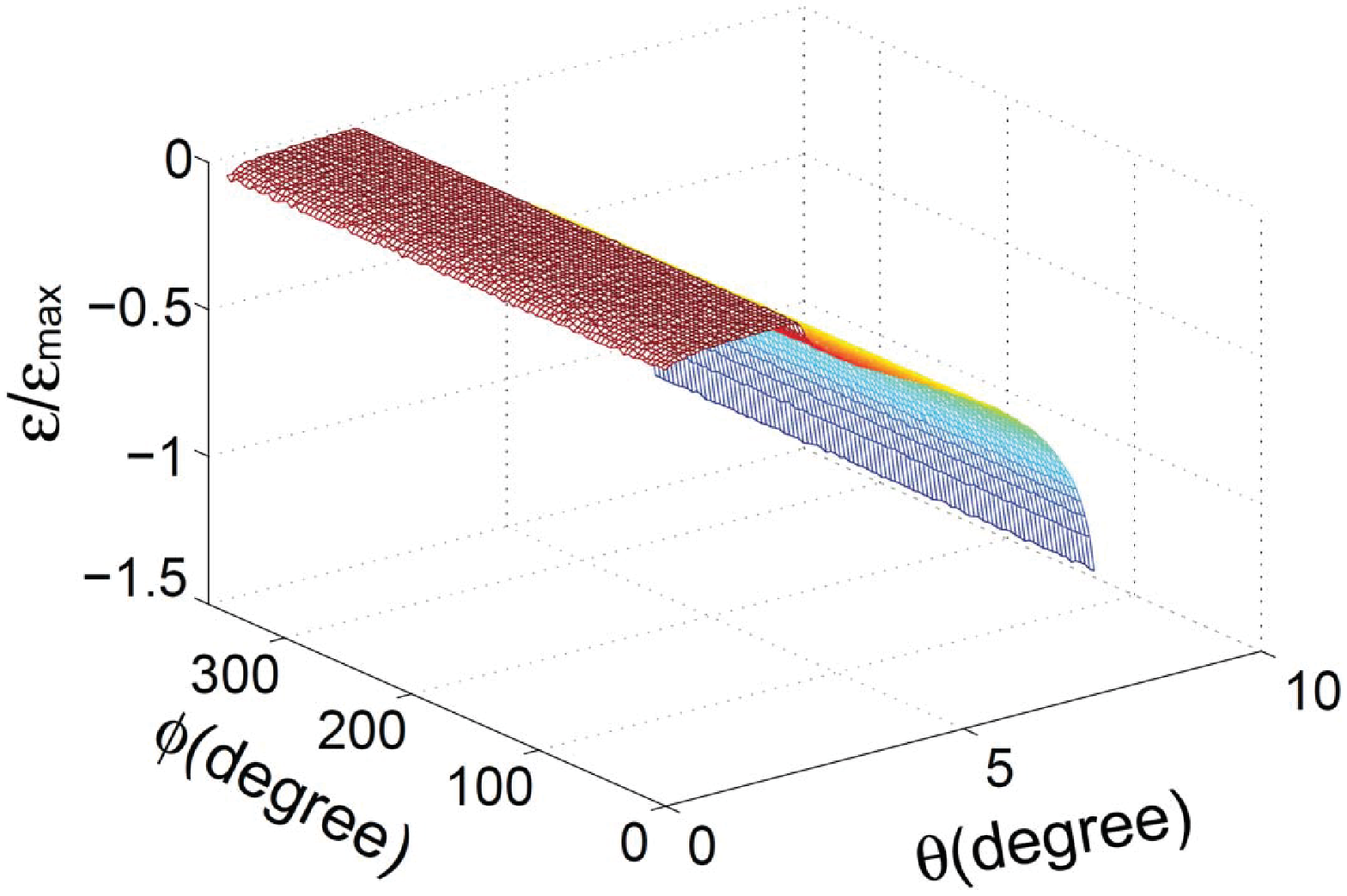} &
\includegraphics[scale=0.28,trim=45 265 45 138,clip]{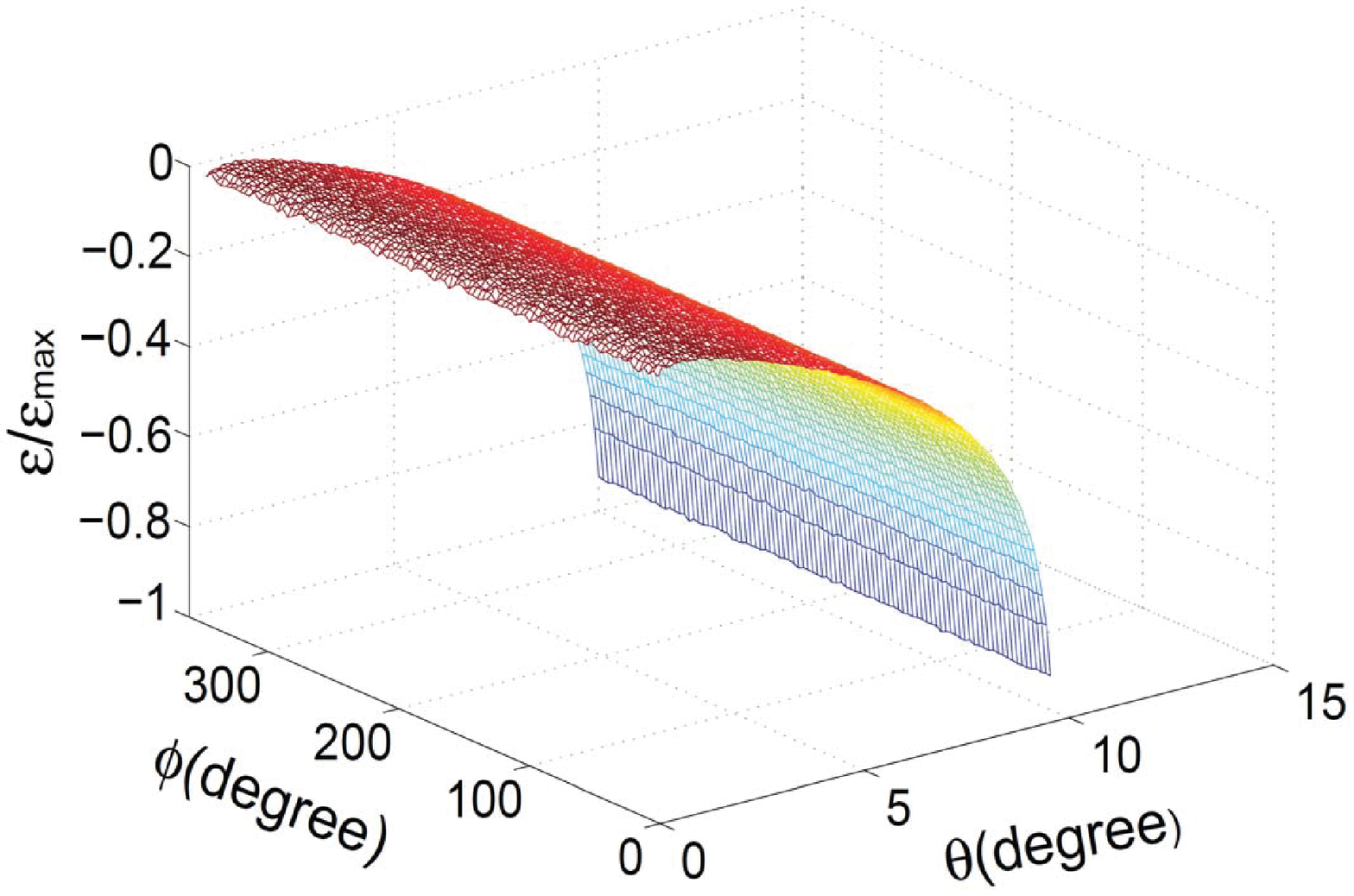} &
\includegraphics[scale=0.28,trim=45 265 45 138,clip]{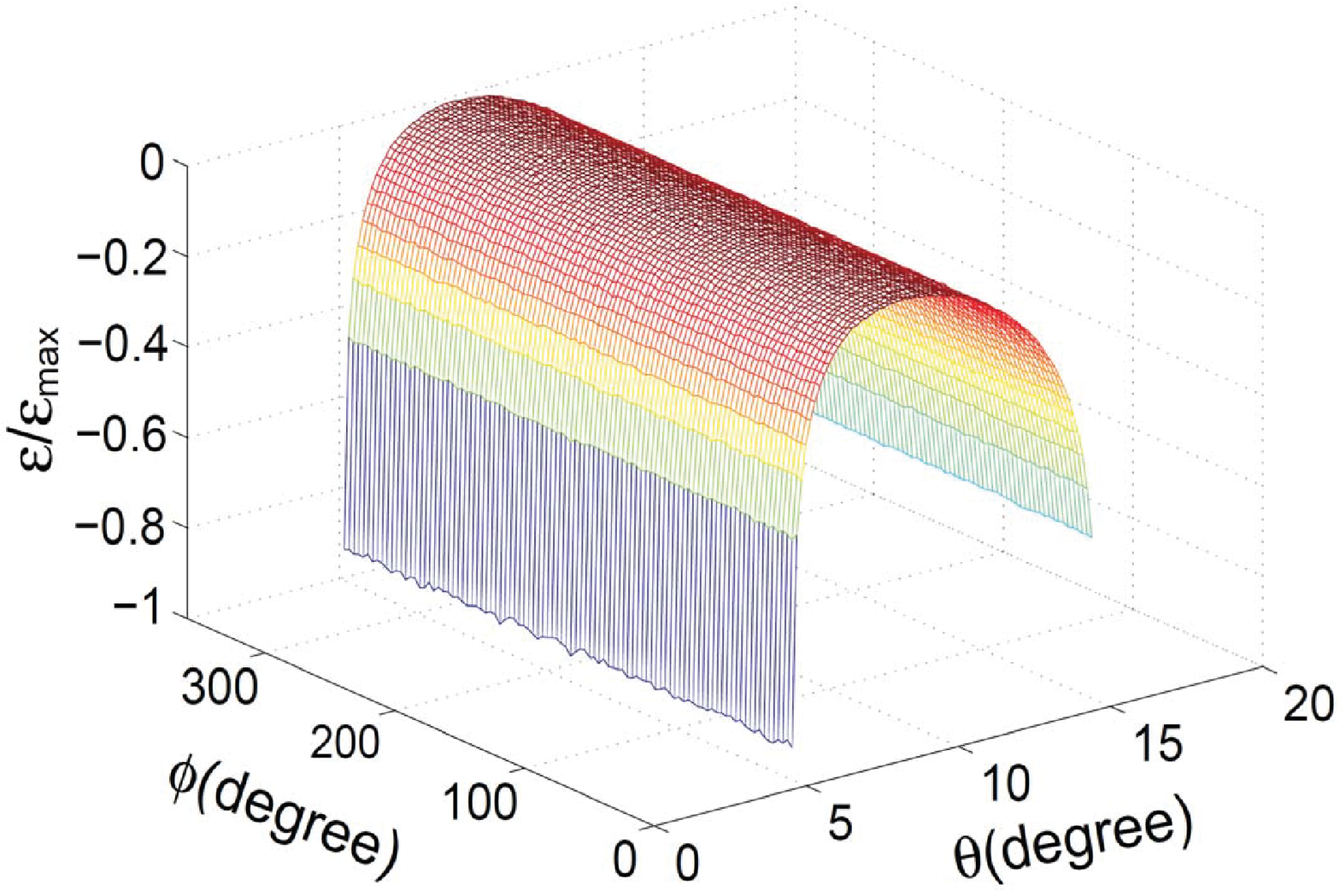} \\
\includegraphics[scale=0.21,trim=0 0 0 0,clip]{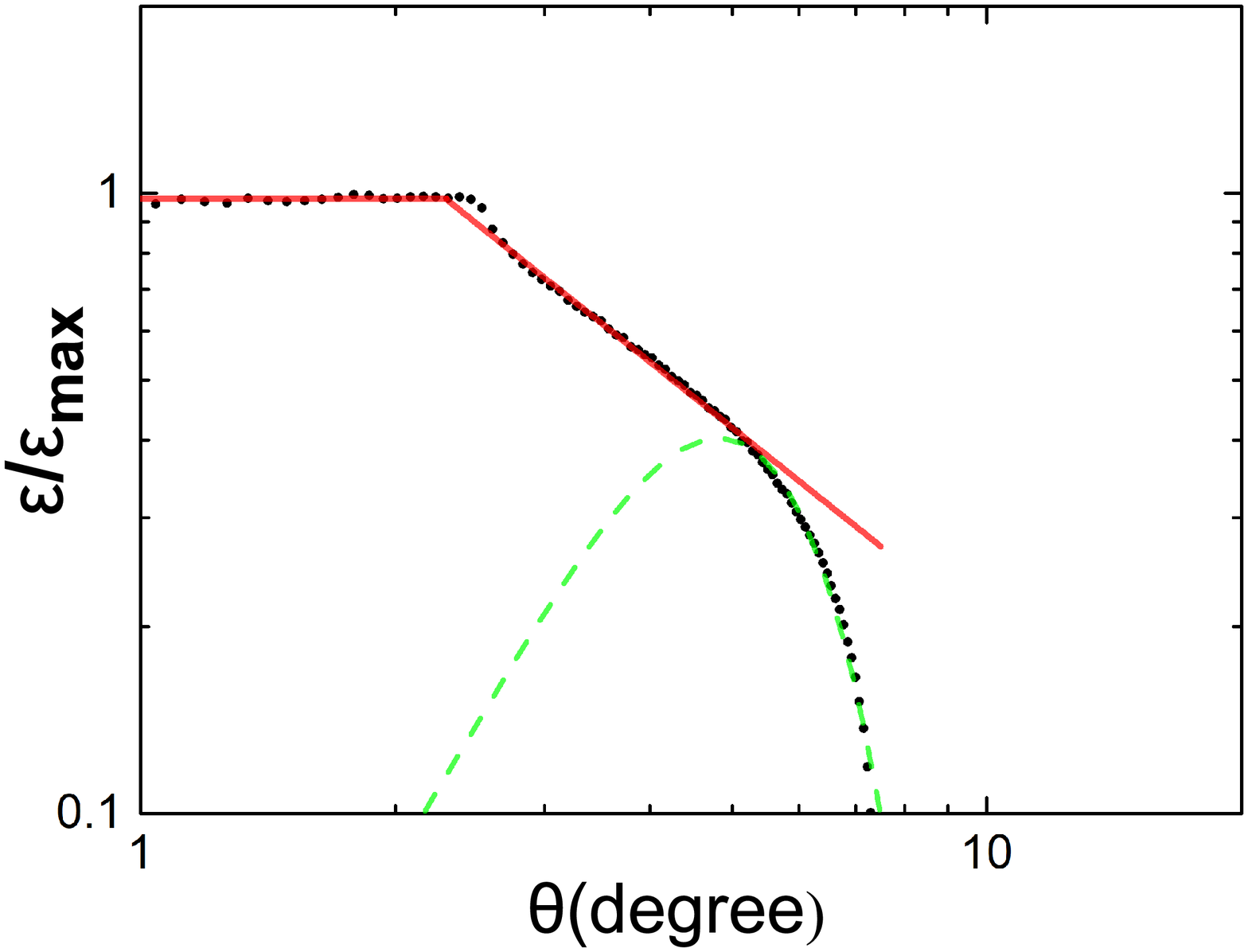} &
\includegraphics[scale=0.21,trim=0 0 0 0,clip]{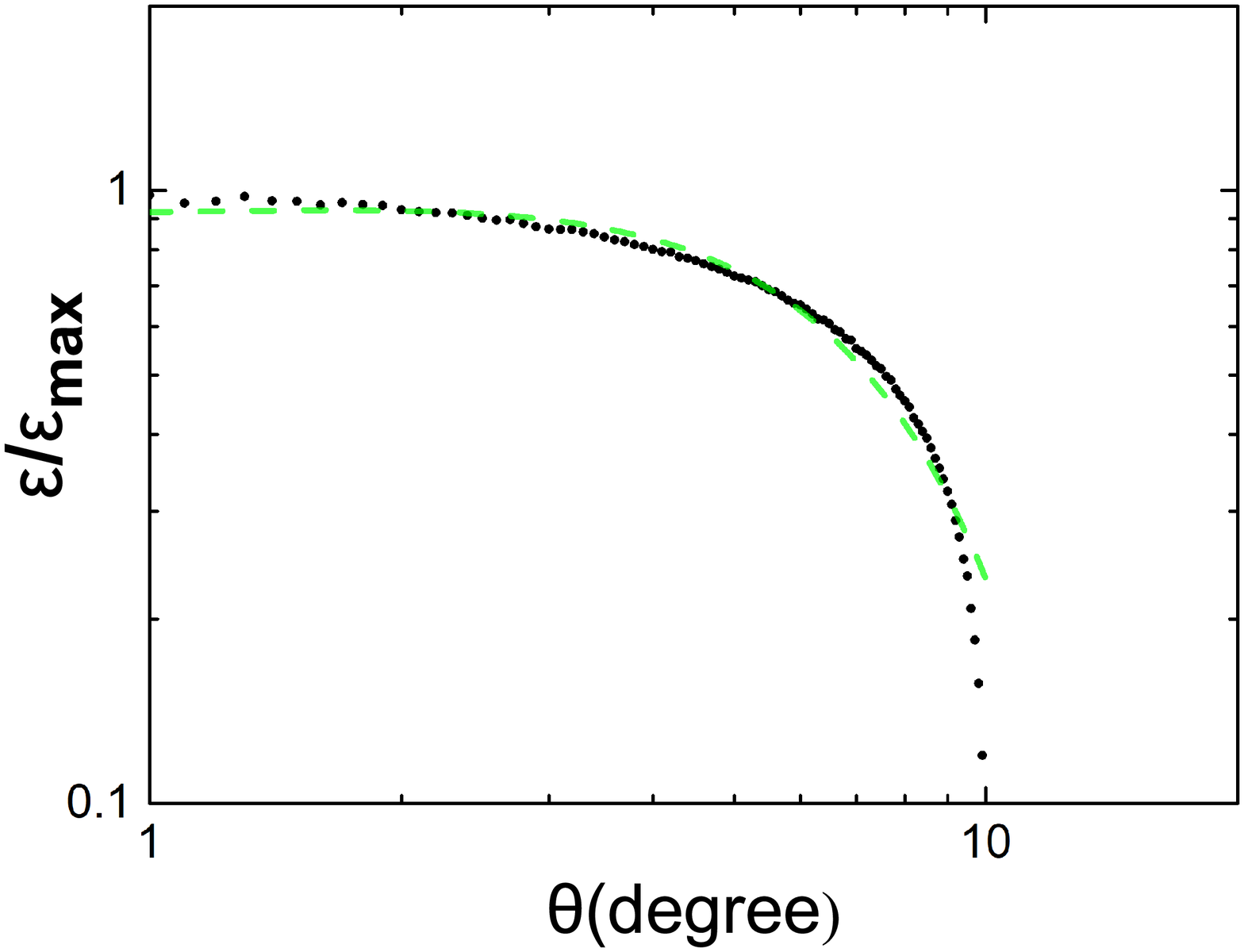} &
\includegraphics[scale=0.21,trim=0 0 0 0,clip]{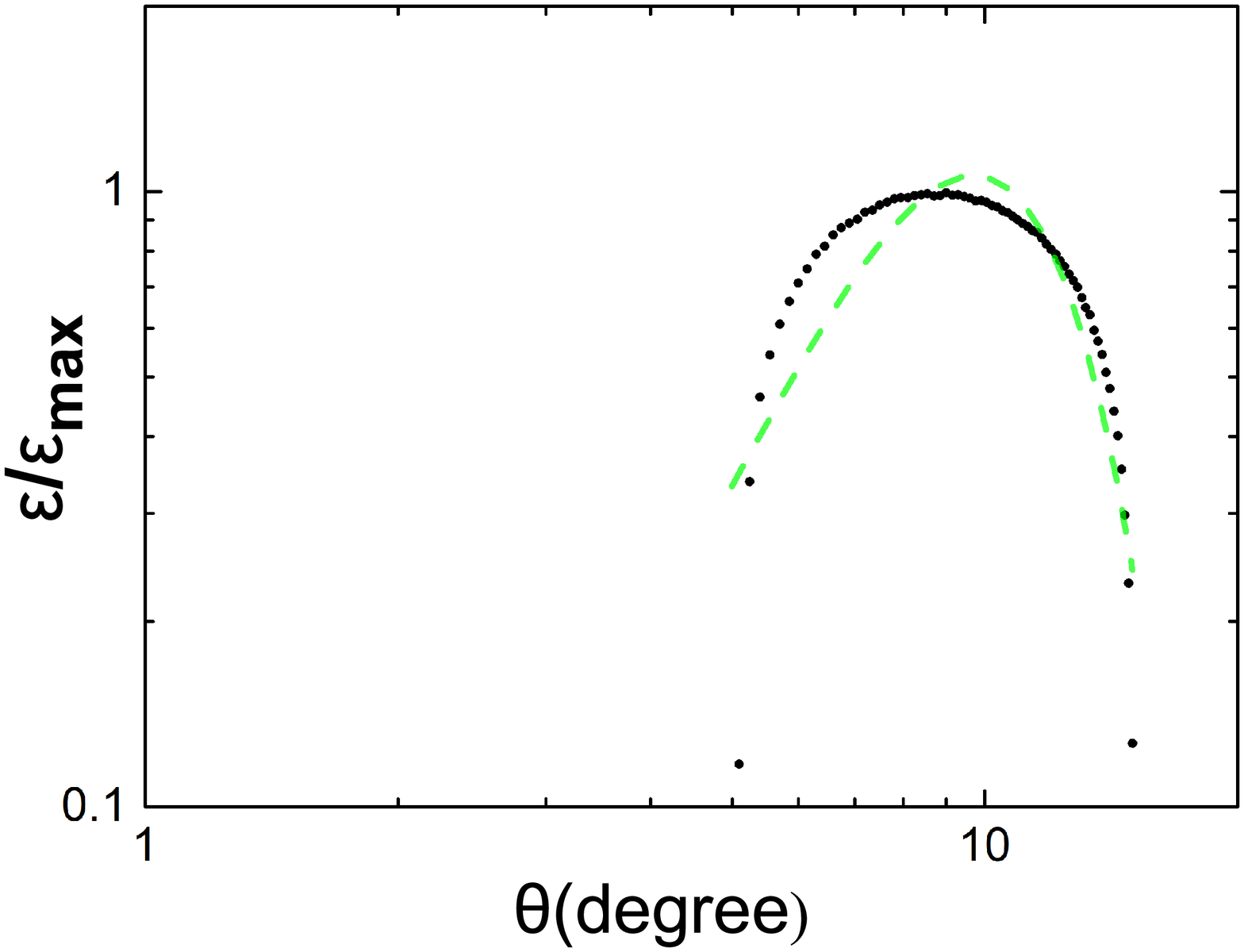} \\

\hline
\end{tabular}}
\caption{Distribution of energy density $\varepsilon$ for GRBs with a precessing jet.
From left to right panel, the value of $\theta_{\rm pre}=2.5^{\circ}$, $5^{\circ}$, and $10^{\circ}$ are adopted, respectively. The dependence of $\varepsilon$ on $(x,y)$, $(\theta, \phi)$, and $\theta$ are shown in the upper,
middle, and lower sub-figures in each panel, respectively.}\label{myFigB}
\end{figure*}

\begin{figure*}\centering
\scriptsize{
\begin{tabular}{ccc}
$\theta_{\rm pre}=2.5^\circ$ & $\theta_{\rm pre}=5^\circ$ & $\theta_{\rm pre}=10^\circ$ \\
\includegraphics[width=.30\textwidth]{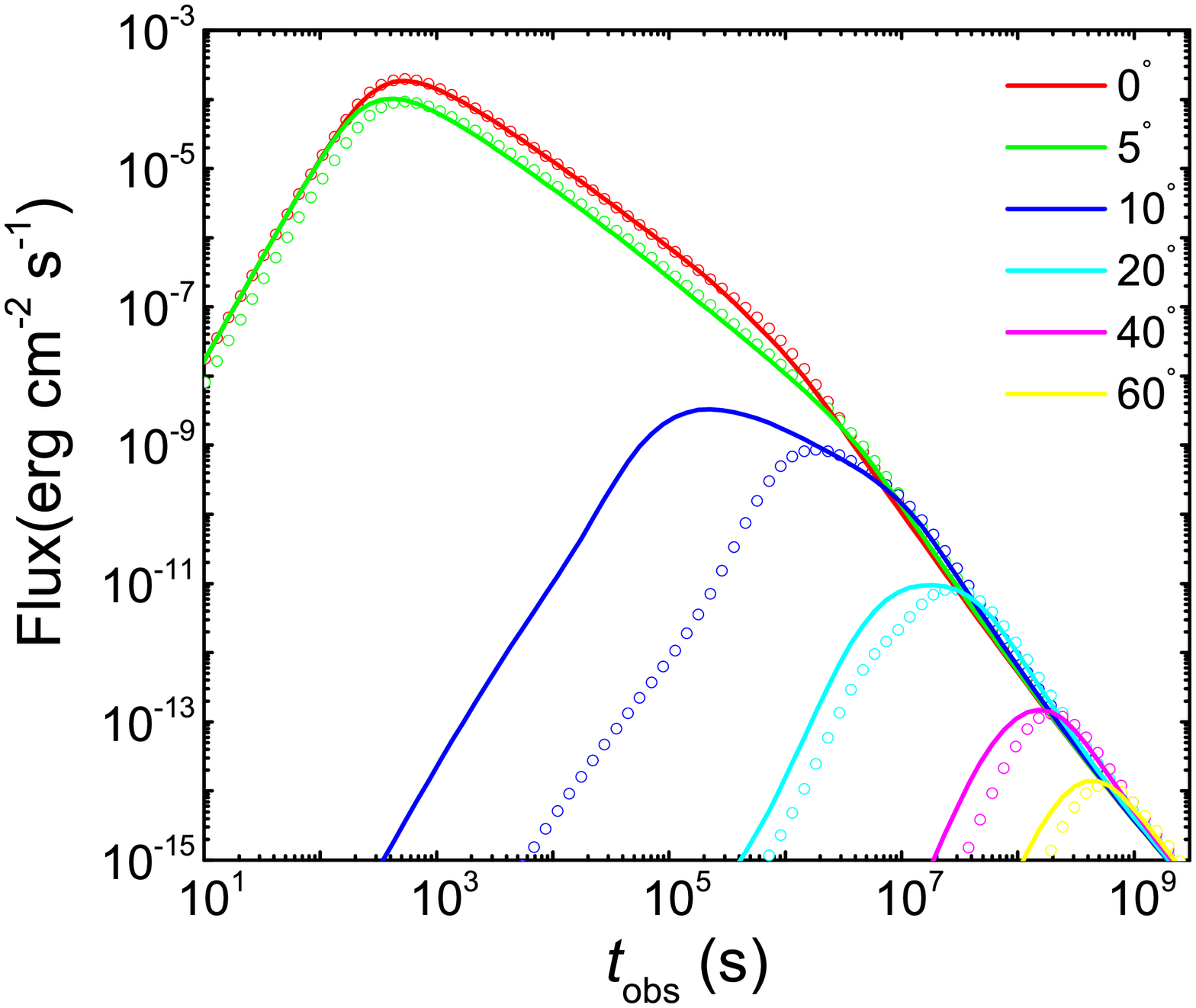} &
\includegraphics[width=.30\textwidth]{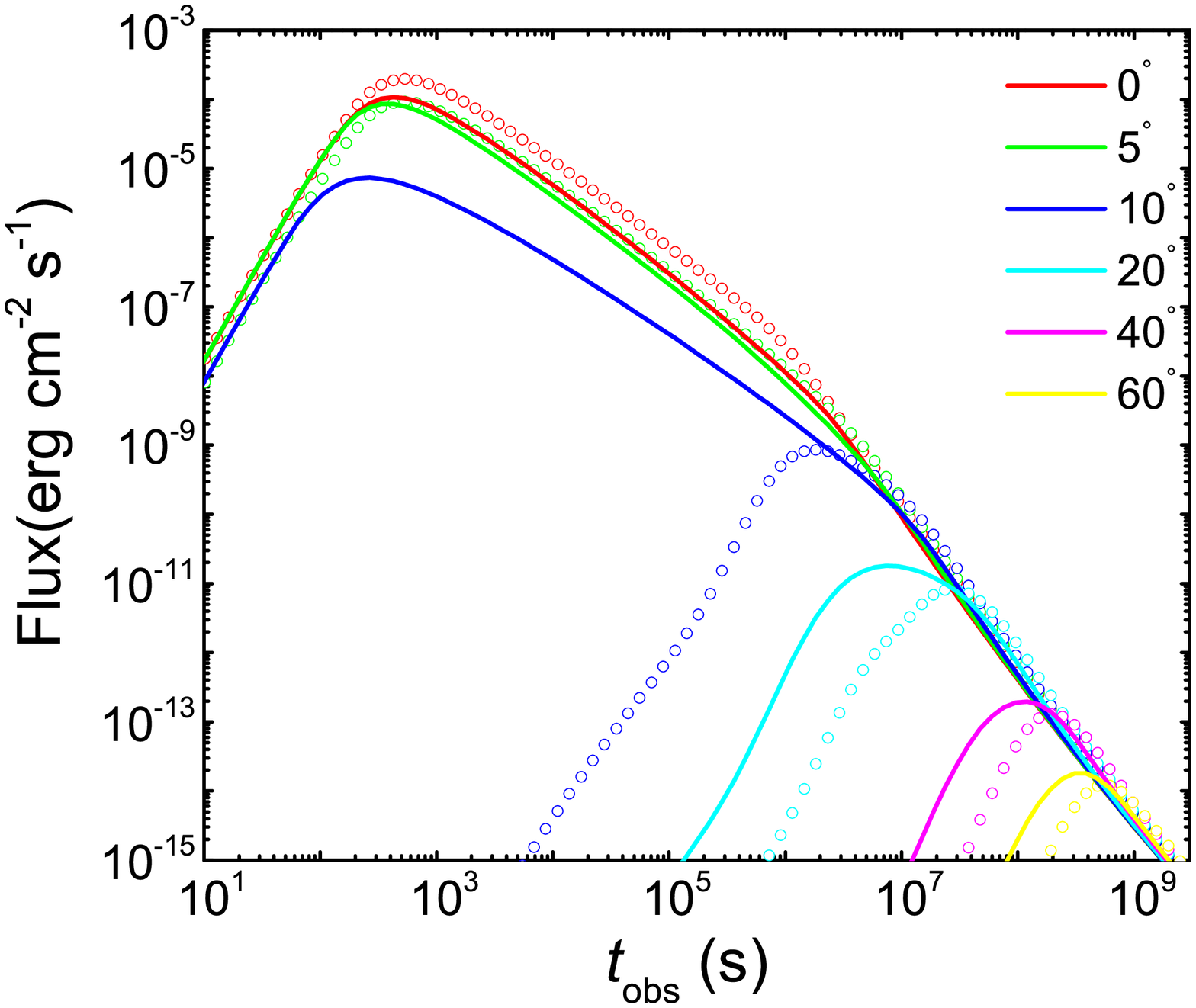} &
\includegraphics[width=.30\textwidth]{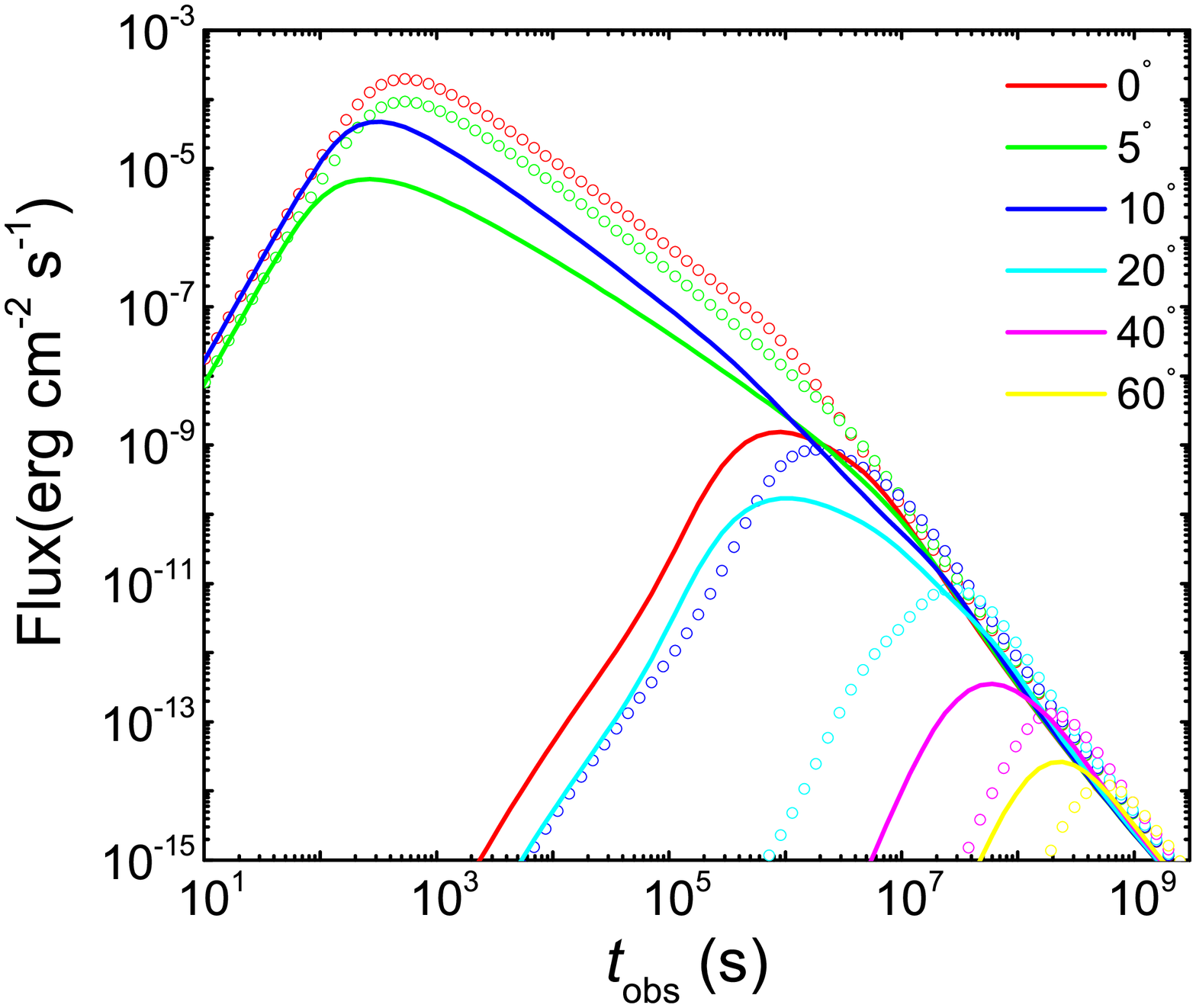} \\
\end{tabular}}
\caption{The X-ray afterglows for GRBs with a precessing jet and different $\theta_{\rm pre}$,
where the viewing angle $\theta_{\rm v}=0^{\rm \circ}$ (red solid line), $5^{\rm \circ}$ (green solid line), $10^{\rm \circ}$ (blue solid line), $20^{\rm \circ}$ (cyan solid line), $40^{\rm \circ}$ (magenta solid line), and $60^{\rm \circ}$ (yellow solid line) are adopted.
The ``$\circ$'' represents the X-ray afterglows from the situation with $\theta_{\rm pre}=0^{\circ}$ and the afterglows with the same viewing angle are plotted with the same color.}
\label{myFigC}
\end{figure*}

\begin{figure*}\centering
\centering\scriptsize{
\begin{tabular}{|c|c|c|}
\hline
 &  &  \\
$\theta_{\rm pre}=2.5^\circ, t_{\rm d}=0.1\tau$ & $\theta_{\rm pre}=5^\circ, t_{\rm d}=0.1\tau$ & $\theta_{\rm pre}=10^\circ, t_{\rm d}=0.1\tau$ \\
\includegraphics[scale=0.29,trim=54 280 50 160,clip]{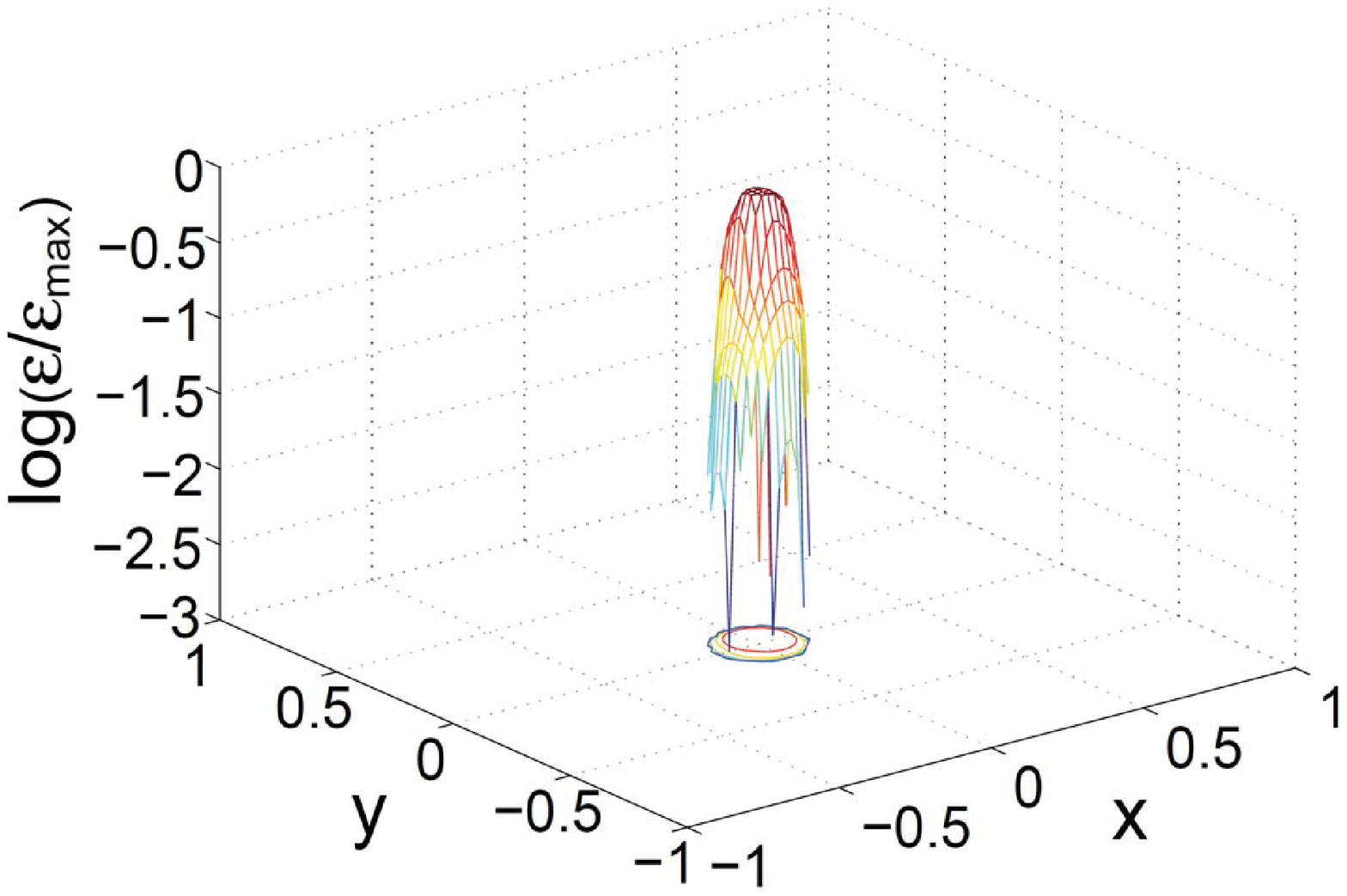} &
\includegraphics[scale=0.29,trim=54 280 50 160,clip]{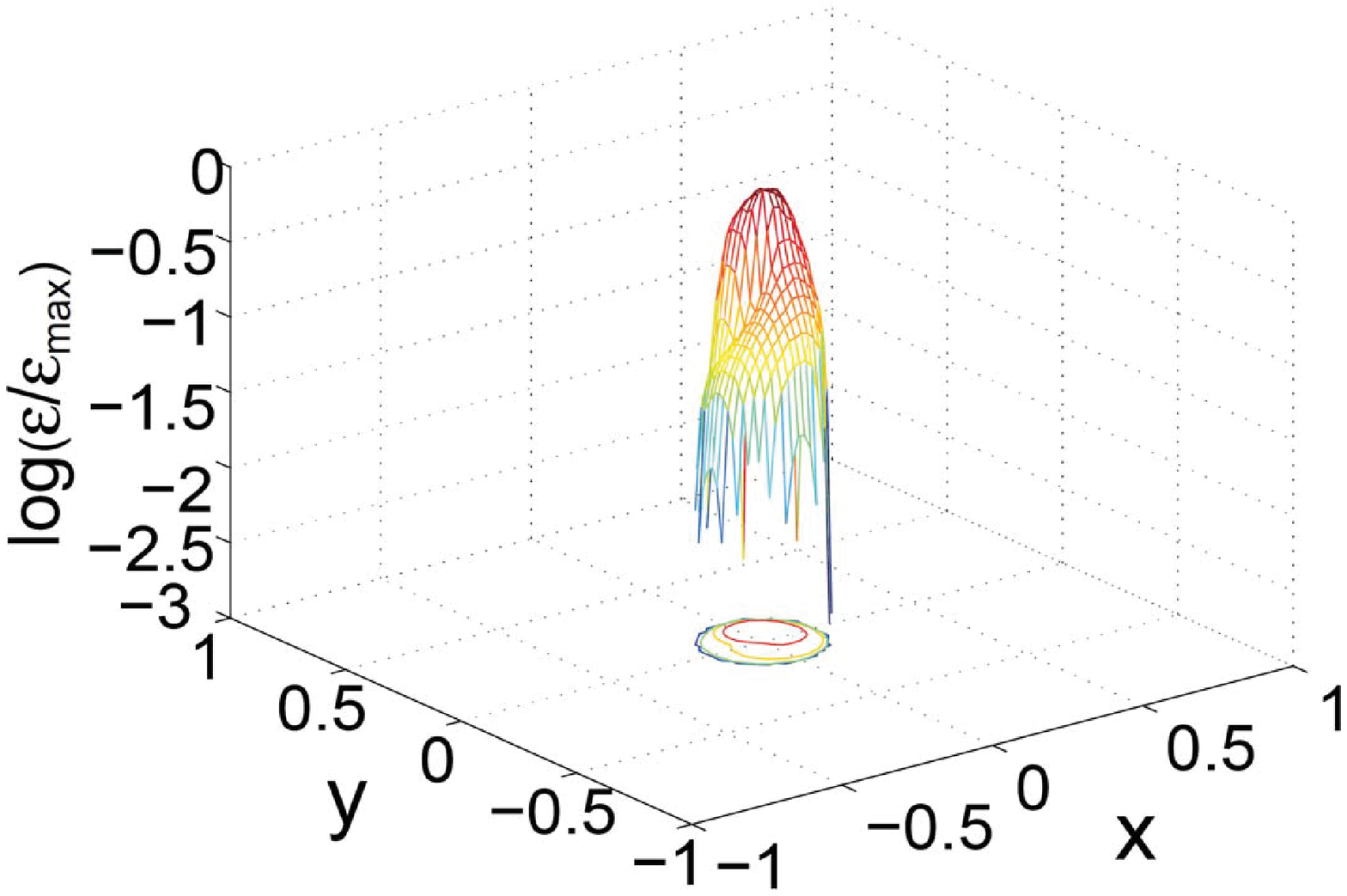} &
\includegraphics[scale=0.29,trim=54 280 50 160,clip]{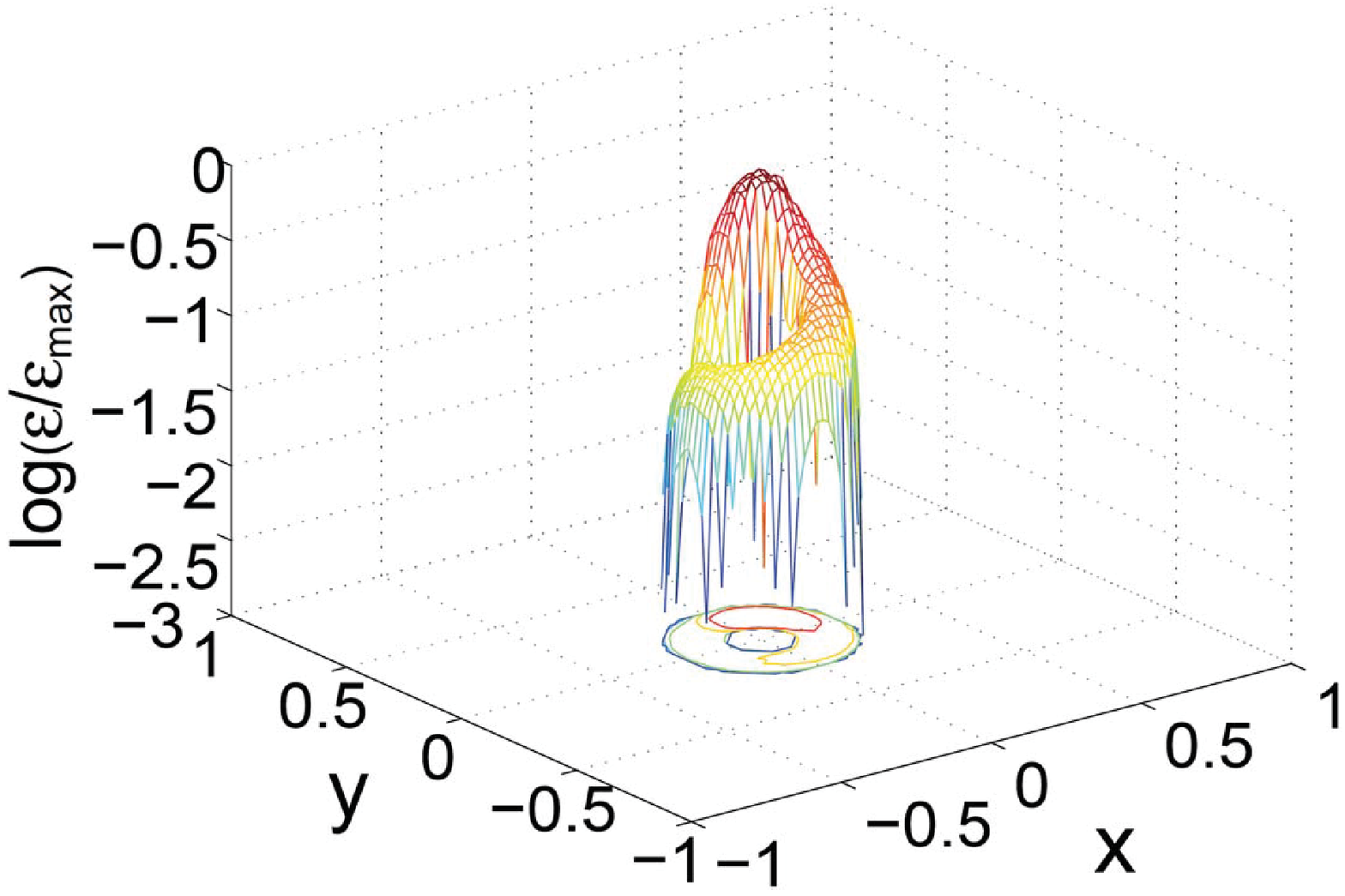} \\
$\theta_{\rm pre}=2.5^\circ, t_{\rm d}=\tau$ & $\theta_{\rm pre}=5^\circ, t_{\rm d}=\tau$ & $\theta_{\rm pre}=10^\circ, t_{\rm d}=\tau$ \\
\includegraphics[scale=0.29,trim=54 280 50 160,clip]{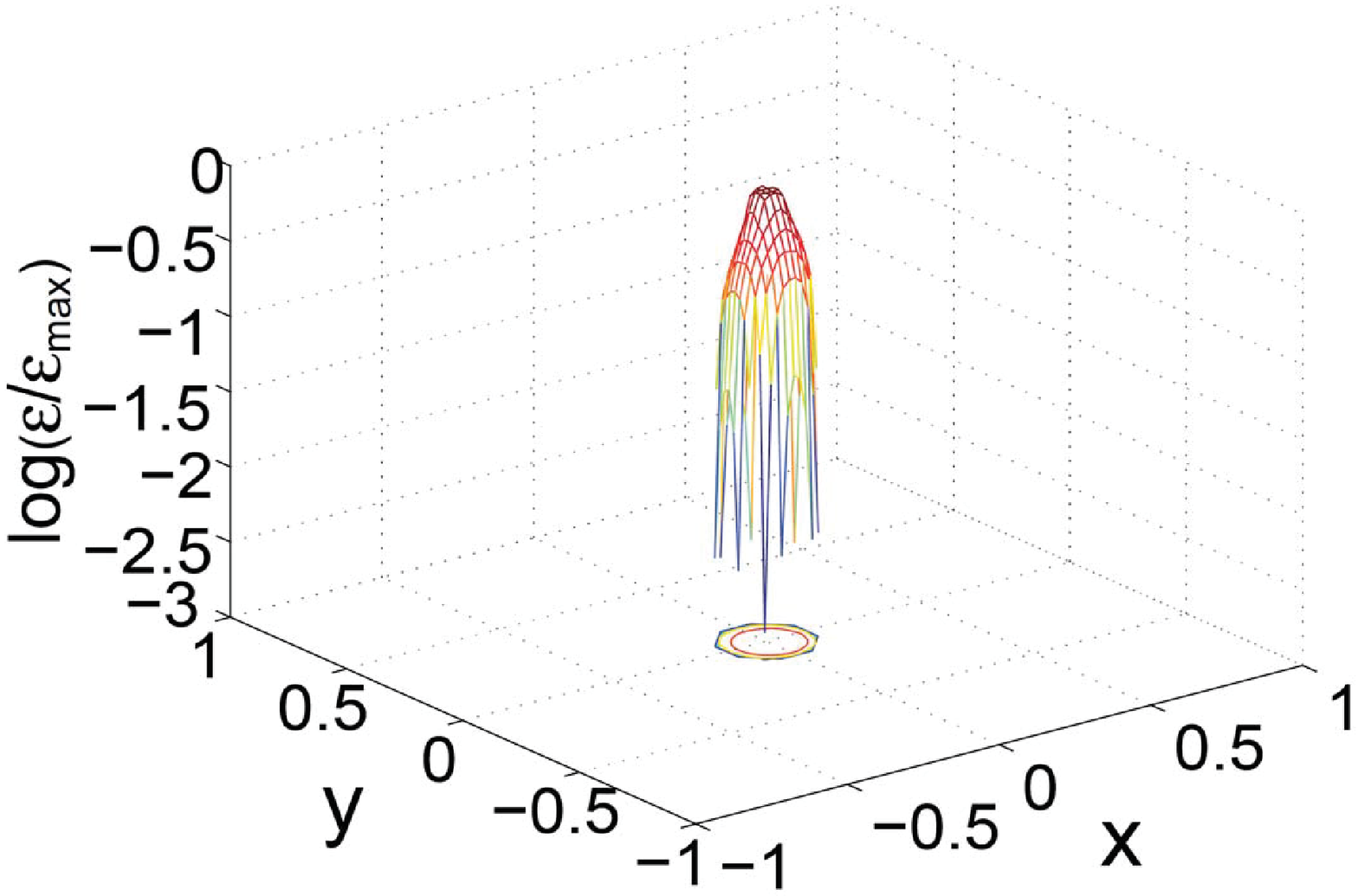} &
\includegraphics[scale=0.29,trim=54 280 50 160,clip]{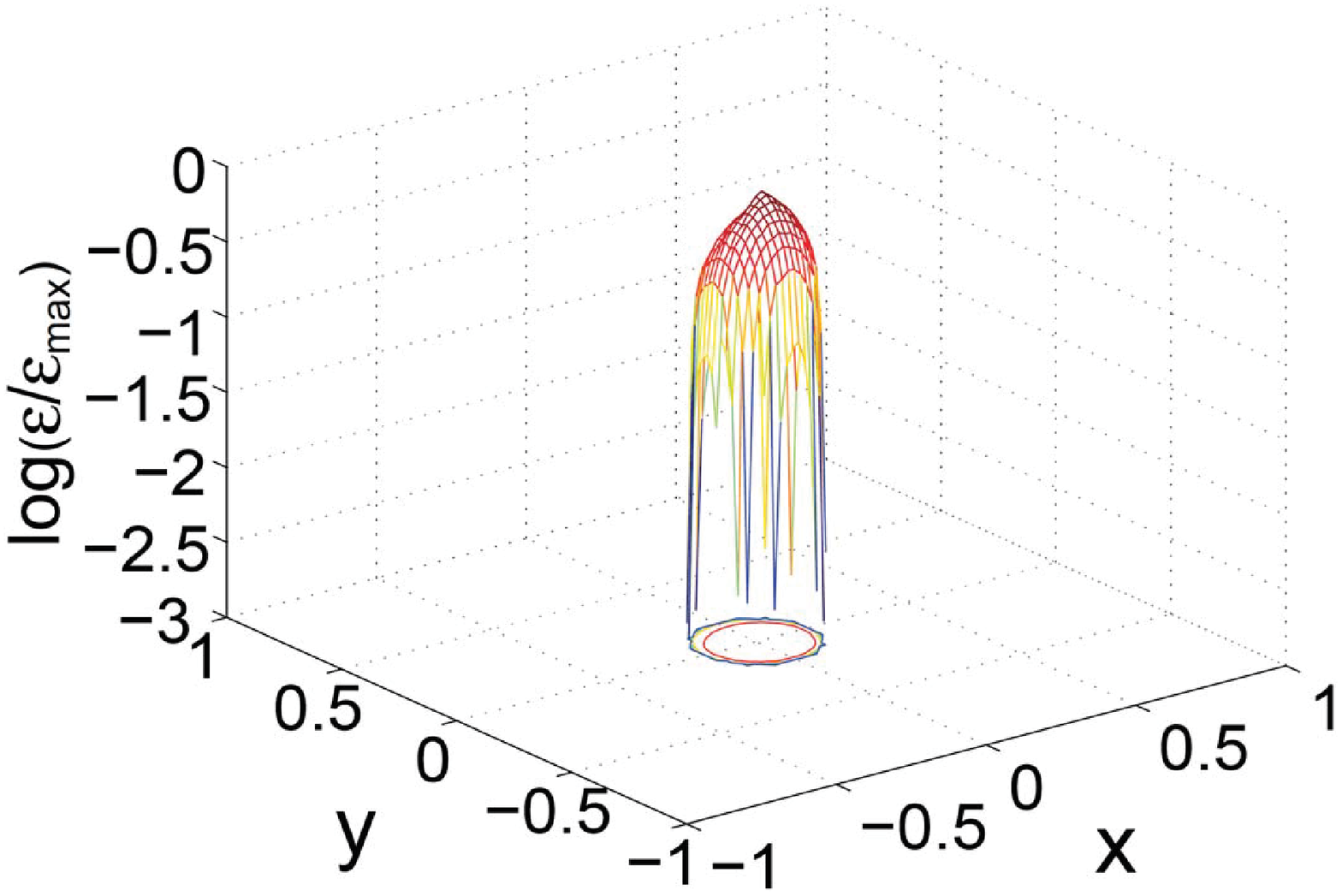} &
\includegraphics[scale=0.29,trim=54 280 50 160,clip]{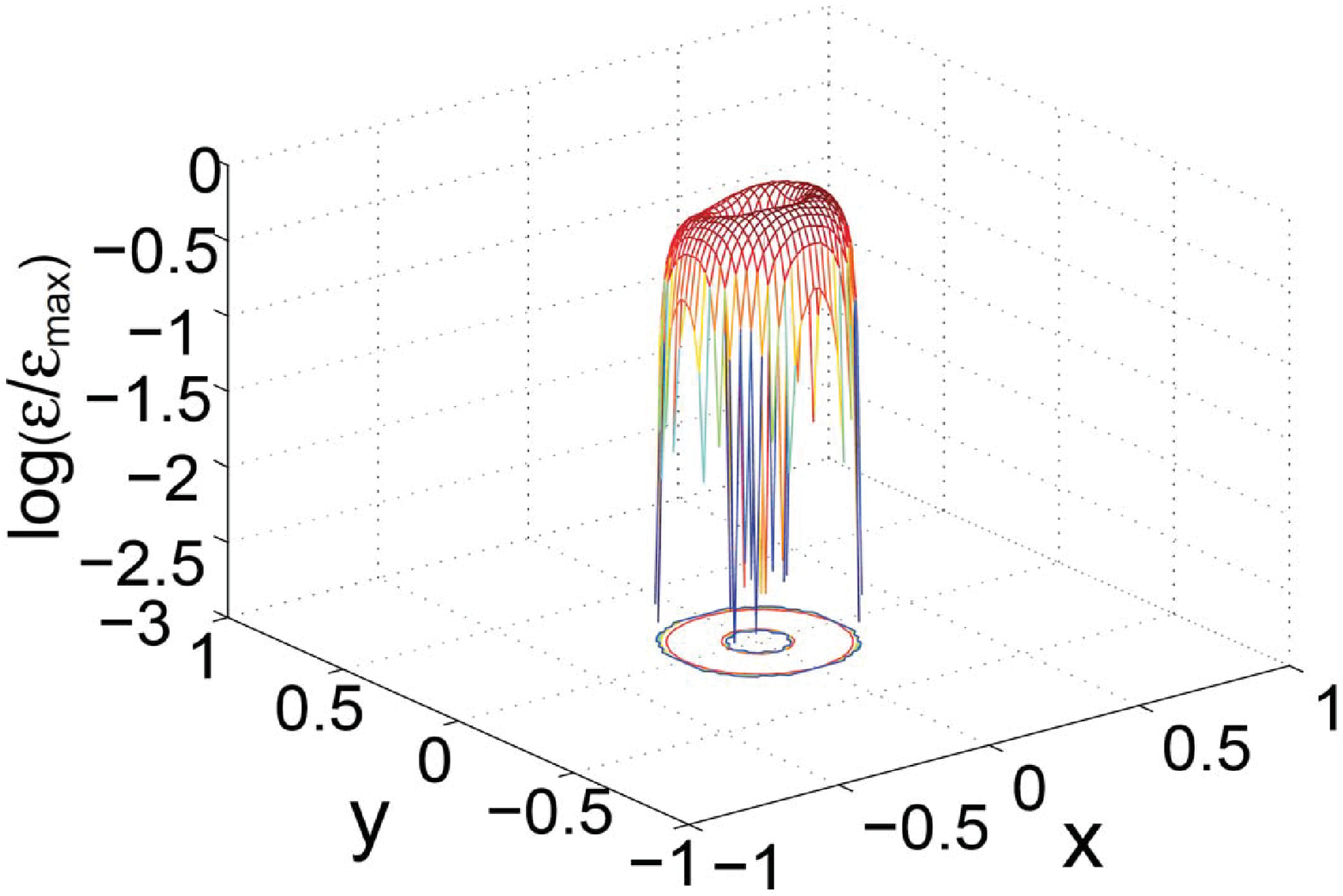} \\
\hline
\end{tabular}}
\caption{Dependence of $\varepsilon$ on $(x,y)$
for a precessing jet with an evolving $P(t)$.}
\label{myFigD}

\end{figure*}
\begin{figure*}\centering
\centering
\begin{tabular}{|c|c|c|}
\hline
$\theta_{\rm pre}=2.5^\circ, t_{\rm d}=0.1\tau$ & $\theta_{\rm pre}=5^\circ, t_{\rm d}=0.1\tau$ & $\theta_{\rm pre}=10^\circ, t_{\rm d}=0.1\tau$ \\
\includegraphics[width=.29\textwidth]{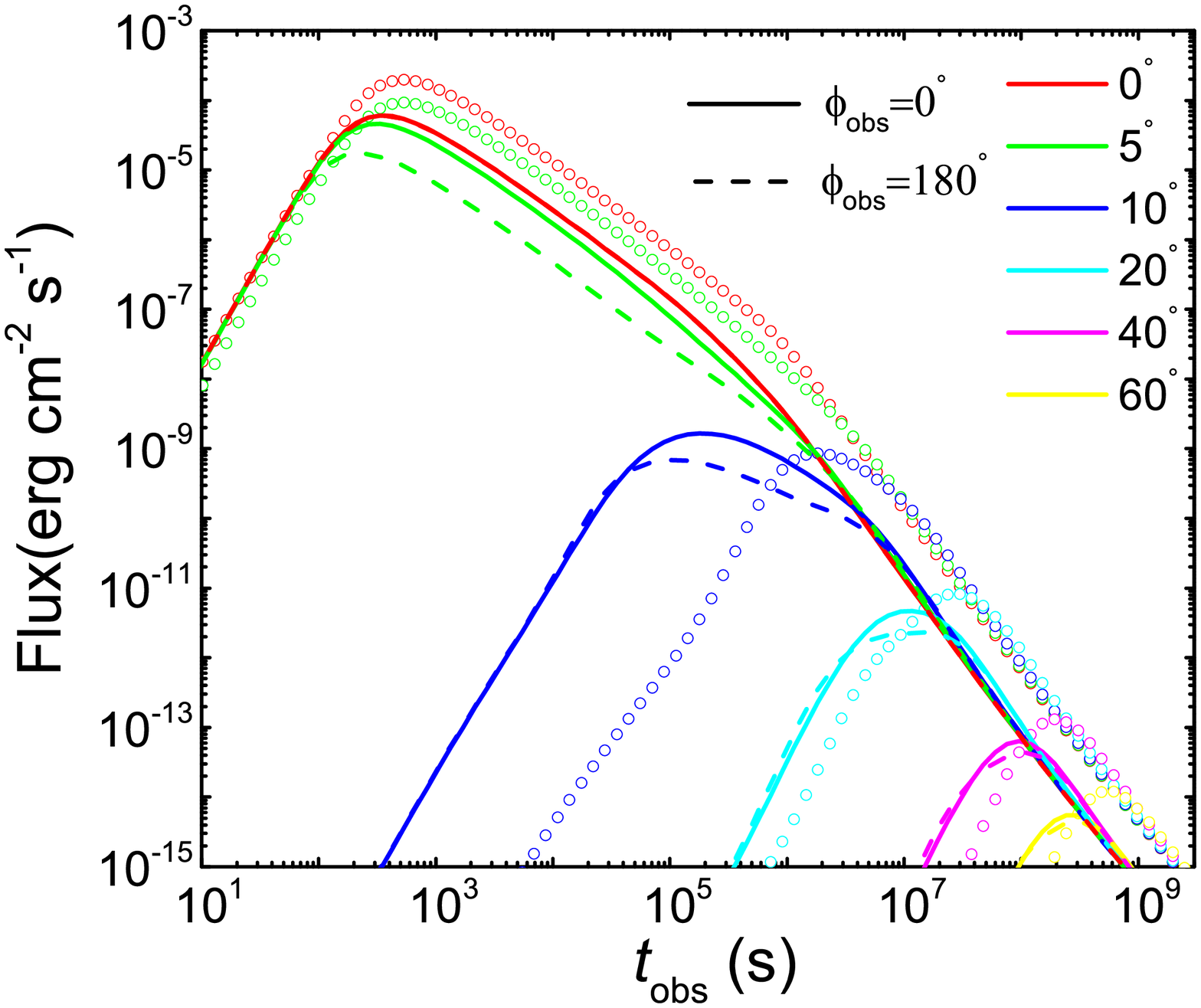} &
\includegraphics[width=.29\textwidth]{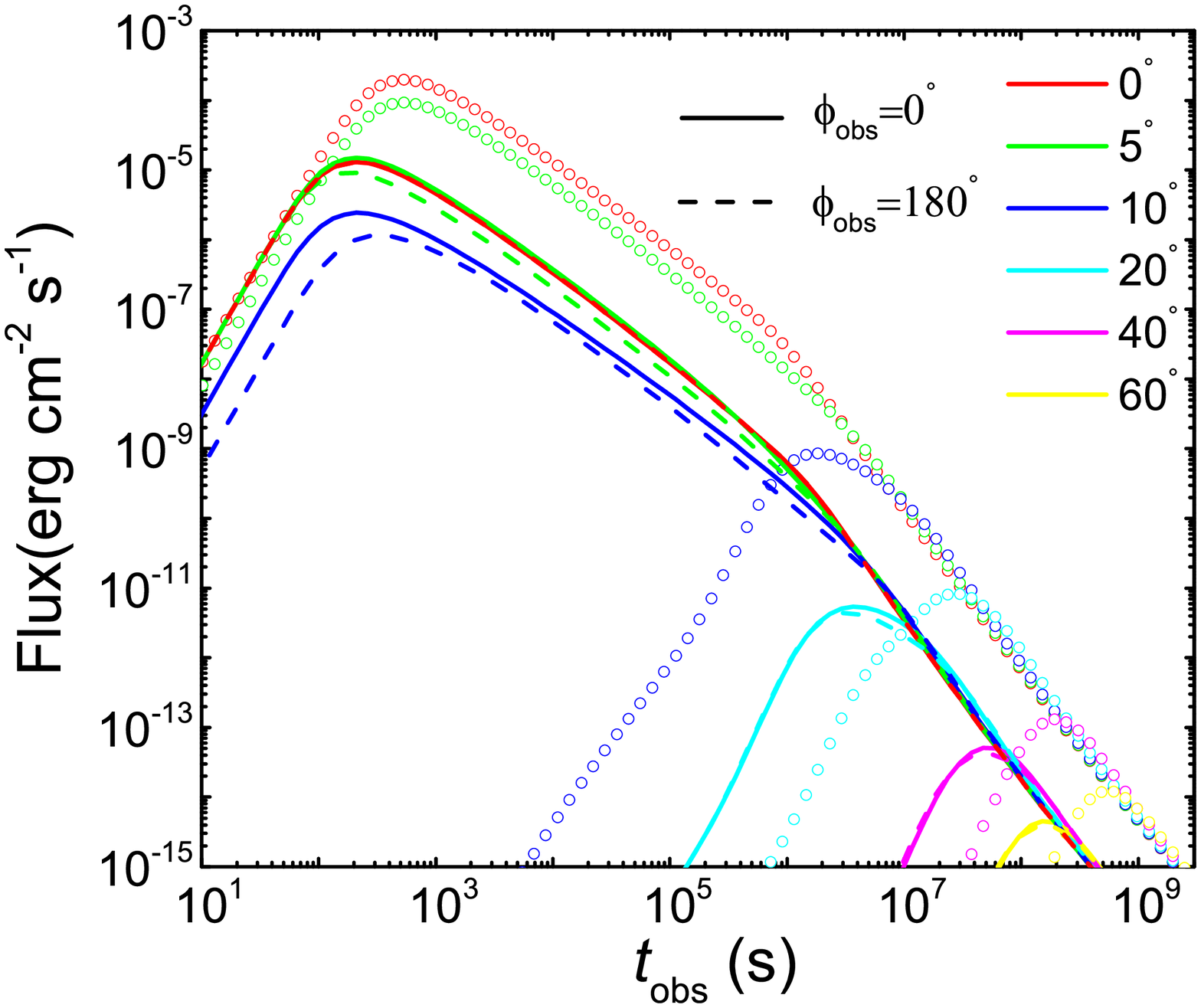} &
\includegraphics[width=.29\textwidth]{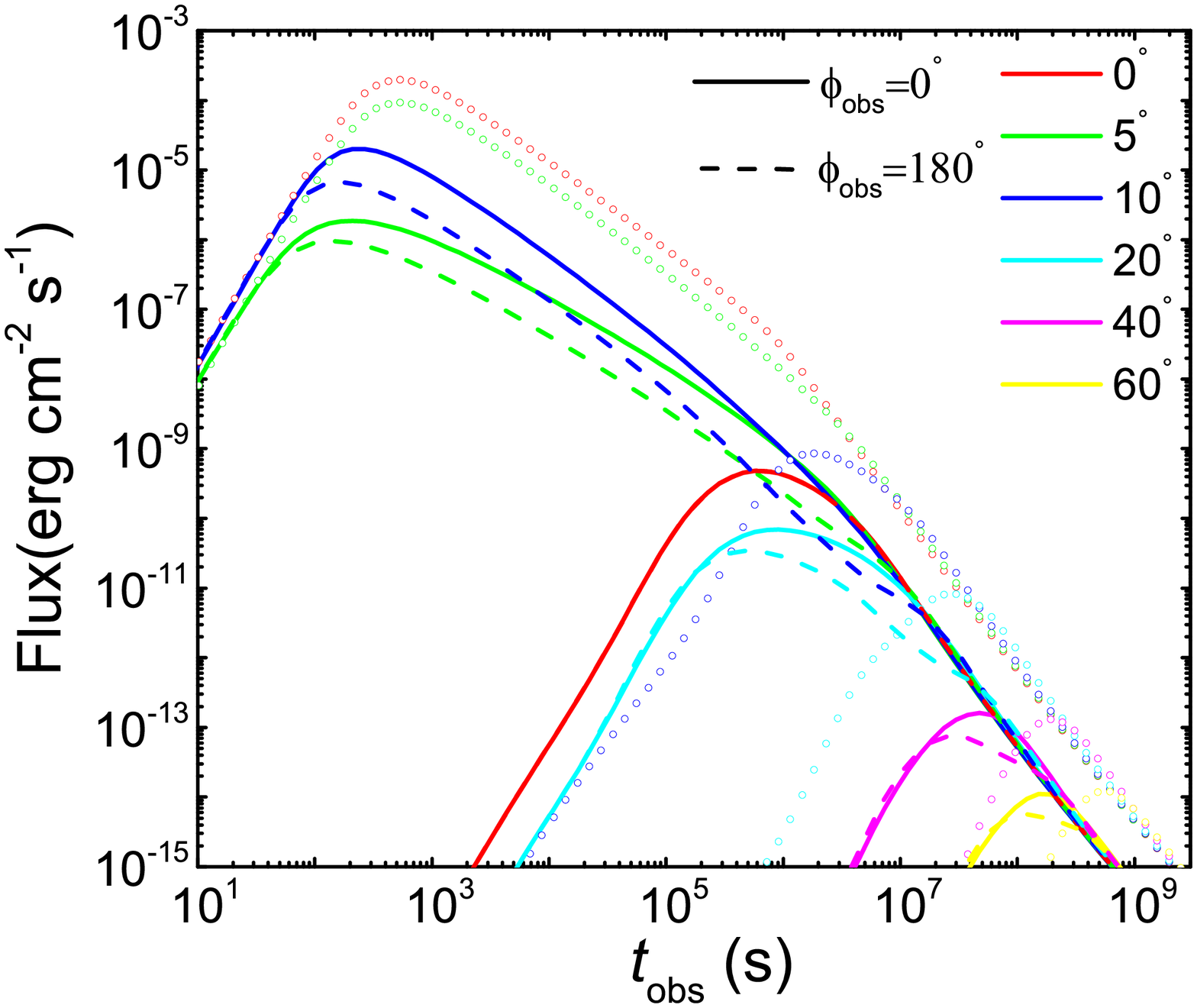}\\
$\theta_{\rm pre}=2.5^\circ, t_{\rm d}=\tau$ & $\theta_{\rm pre}=5^\circ, t_{\rm d}=\tau$ & $\theta_{\rm pre}=10^\circ, t_{\rm d}=\tau$ \\
\includegraphics[width=.29\textwidth]{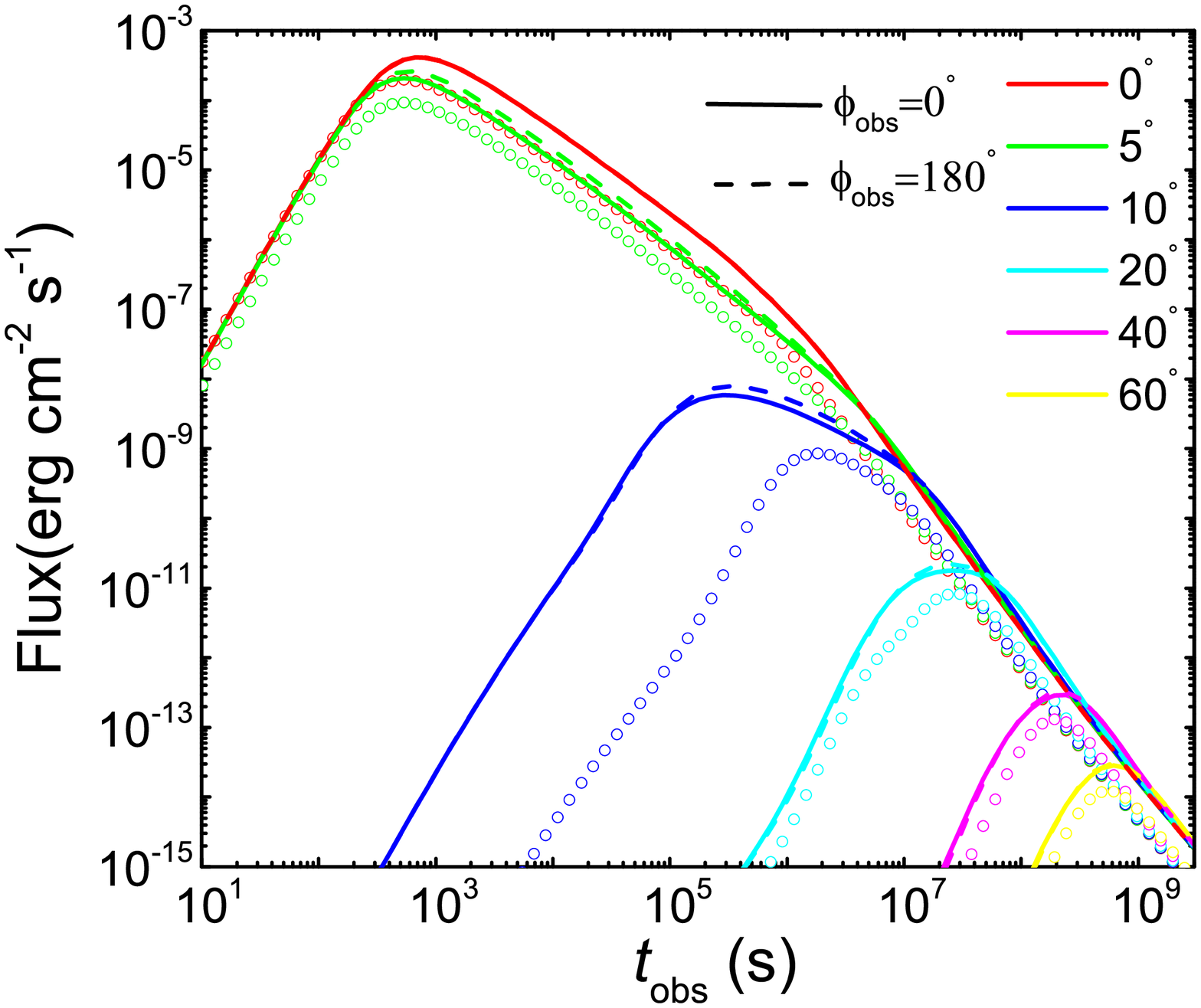}&
\includegraphics[width=.29\textwidth]{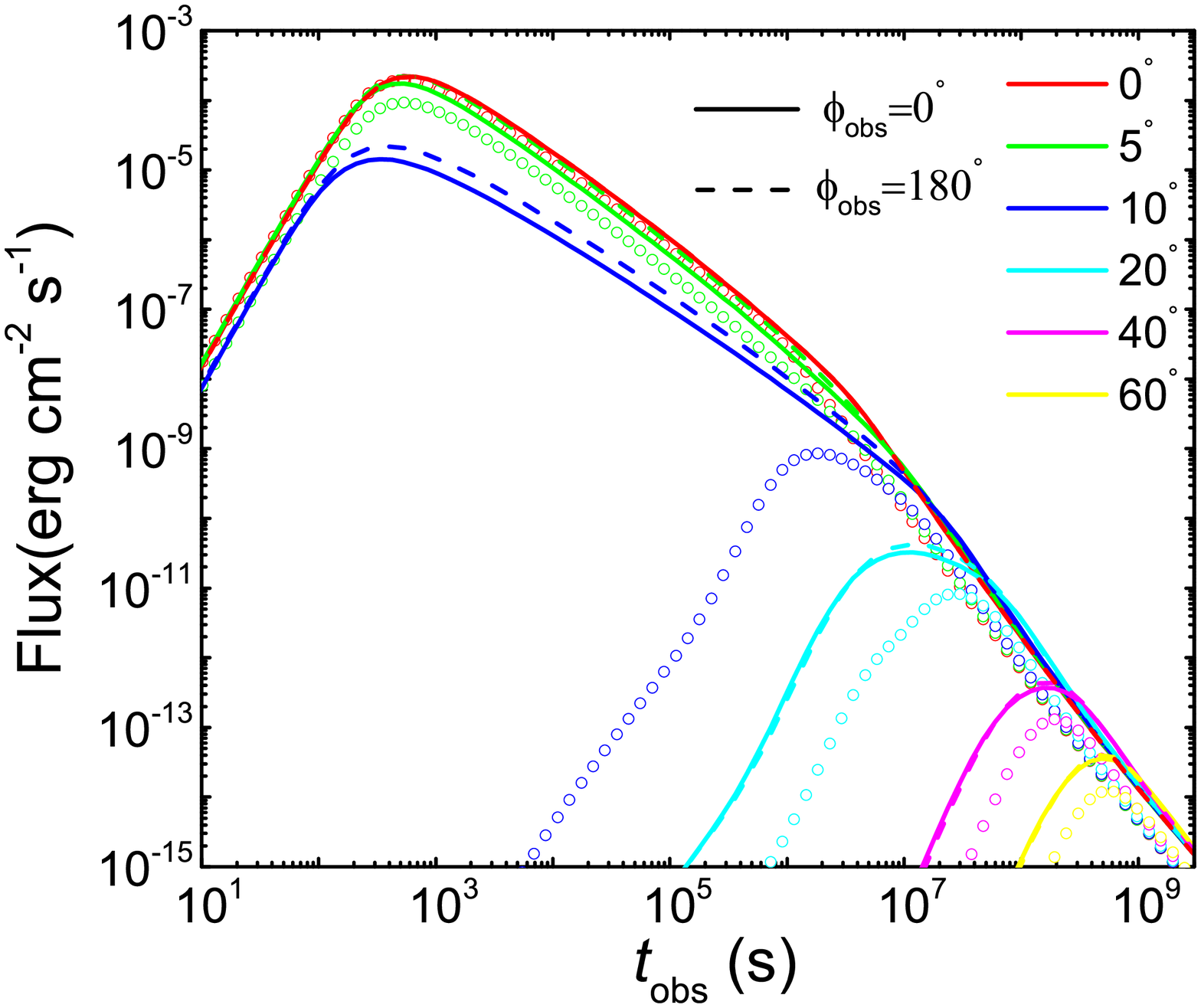}&
\includegraphics[width=.29\textwidth]{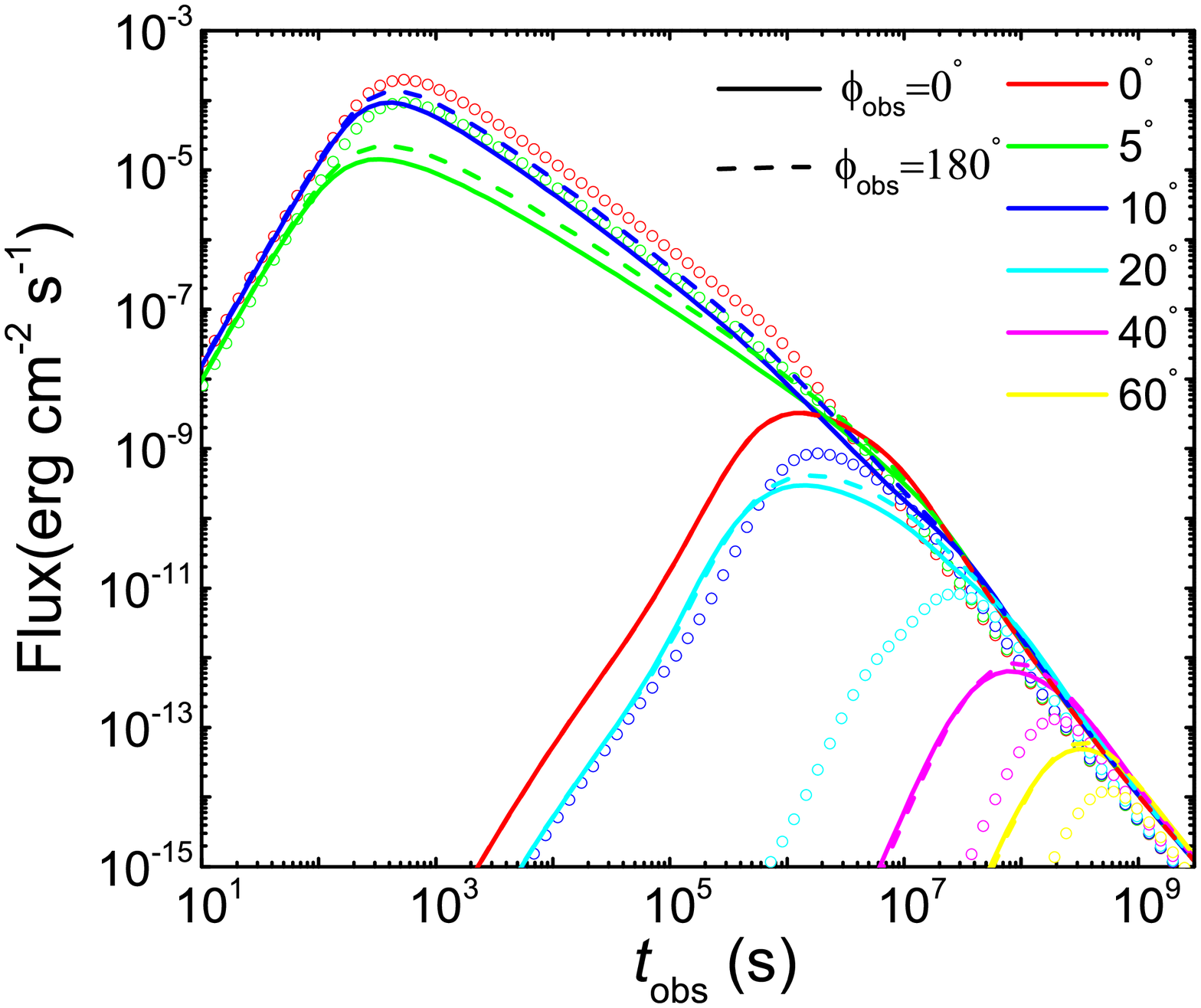}\\
\hline
\end{tabular}
\caption{The X-ray afterglows for an evolving jet with $t_{\rm d}=0.1\tau$ (upper sub-figures) and $t_{d}=\tau$ (lower sub-figures),respectively.
The solid and dashed lines
represent the results for an observer with viewing angle of
$(\theta_{\rm obs},\phi_{\rm obs})=(\theta_{\rm obs},0^{\circ})$ and $(\theta_{\rm obs},180^{\circ})$, respectively.
Here, $\theta_{\rm obs}=0^{\circ},5^{\circ},10^{\circ},20^{\circ},40^{\circ},60^{\circ}$ are studied in each sub-figure and the meaning of ``$\circ$'' symbol is the same as that in Figure~\ref{myFigC}.
}\label{myFigE}
\end{figure*}

We first study the jet structure with $P(t)=\rm constant$, $t_{\rm end}=\tau$, and different $\theta_{\rm pre}$.
The obtained distribution of energy density $\varepsilon$ per solid angle is shown in Figure \ref{myFigB},
where $\varepsilon_{\max}$ is the maximum energy density and
$\theta_{\rm pre}=2.5^{\circ}$ (left panel),
$5^{\circ}$ (middle panel), and $10^{\circ}$ (right panel) are adopted, respectively.
In this figure, the upper sub-figures in each panel
present the dependence of $\varepsilon$ on $x=\sin\theta\cos\phi$ and $y=\sin\theta\sin\phi$.
For a given $\theta$ in these sub-figures, i.e., $x^2+y^2=\rm constant$,
the value of $\varepsilon$ is the same for different $\phi$.
This can be easily found in the middle sub-figures of each panel,
which show the dependence of $\varepsilon$ on $\theta$ and $\phi$.
In the situation of $\theta_{\rm open}>\theta_{\rm pre}$ (left panel),
one can find that $\varepsilon$ remains constant in the region with $\theta<\theta_{\rm open}-\theta_{\rm pre}$
and follows a power-law decay in the region of $\theta>\theta_{\rm open}-\theta_{\rm pre}$.
Since the $\varepsilon$ does not evolve in $\phi$-direction,
we adopt the following function to fit the energy distribution:
\begin{eqnarray}\label{Eq:Energy_distribution}
\frac{\varepsilon(\theta)}{\varepsilon_{\rm max}} = \left\{{\begin{array}{*{20}{c}}
1, & {\theta \leq \theta_{\rm c},}\\
({\theta}/\theta_{\rm c})^{-k}, & {\theta_{\rm c} < \theta < \theta_{\rm m},}
\end{array}} \right.
\end{eqnarray}
where $\theta_{\rm c}$ and $\theta_{\rm m}$ are constants.
The result is shown with red solid line in Figure \ref{myFigB},
and can be read as
$\theta_{\rm c}=2.5^{\circ}$, $\theta_{\rm m}=5^{\circ}$, and $k=1.05$.
Since the jets in Figure \ref{myFigB} are calculated with a narrow-uniform-precessing jet,
the Lorentz factor $\Gamma(\theta,\phi)$ of our obtained jet should be the same for different $(\theta,\phi)$, i.e., $\Gamma(\theta,\phi)=\rm constant$.
This is different from the structured jet adopted in \cite{2017ApJ...850L..41X} (see also, e.g.,
\citealp{2001ApJ...552...72D};
\citealp{2002ApJ...571..876Z};
\citealp{2002MNRAS.332..945R};
\citealp{2003ApJ...591.1075K}).
For $\theta>\theta_{\rm m}$ in the left panel, the $\varepsilon$ suffers from a sharp cut-off decay.
Then, the Gaussian function, i.e.,
\begin{eqnarray}
\varepsilon(\theta)\propto \exp[-(\theta-\theta_{0})^2/2\theta_{\rm g}^2],
\end{eqnarray}
is adopted to fit this part and the result (green dashed line) can be read as $\theta_{0}=4.8^{\circ}$ and $\theta_{\rm g}=1.6^{\circ}$. In summary, the structured jet in the left panel can be described as
\begin{eqnarray}
\begin{array}{l}
\frac{{\varepsilon (\theta )}}{{{\varepsilon _{{\rm{max}}}}}} = \\
\left\{ {\begin{array}{*{20}{c}}
{1,}&{\theta  \le {\theta _{\rm{c}}},}\\
{{{(\frac{\theta }{{{\theta _{\rm{c}}}}})}^{ - k}},}&{{\theta _{\rm{c}}} < \theta  \le {\theta _{\rm{m}}},}\\
{{{(\frac{{{\theta _{\rm{m}}}}}{{{\theta _{\rm{c}}}}})}^{ - k}}\exp [\frac{{ - {{(\theta  - {\theta _0})}^2}}}{{2\theta _{\rm{g}}^2}}],}&{{\theta _{\rm{m}}} < \theta  < {\theta _{{\rm{open}}}} + {\theta _{{\rm{jet}}}}.}
\end{array}} \right.
\end{array}
\end{eqnarray}
According to Figure \ref{myFigB}, it can be easily found that
a power-law function could not describe the energy distribution for the other two situations.
In addition, the cut-off behavior is obviously in these two situations.
Then, we adopt the Gaussian function to fit the energy distribution.
The results are shown with green dashed lines, i.e., $(\theta_{0},\theta_{\rm g})=(1.6^{\circ},5.0^{\circ})$ for $\theta_{\rm open}=\theta_{\rm pre}$
and $(\theta_{0},\theta_{\rm g})=(9.7^{\circ},3.1^{\circ})$ for $\theta_{\rm open}<\theta_{\rm pre}$.
One can find that the Gaussian function can better describe the jet structure for the situation of $\theta_{\rm pre}=5^{\circ}$.
However, it fails to describe the jet structure for the situation with $\theta_{\rm pre}=10^{\circ}$.
The energy distribution in the situation of $\theta_{\rm open}<\theta_{\rm pre}$ is a ring-shaped jet,
but not a uniform ring-shaped jet (\citealp{2005ApJ...631.1022G};
\citealp{2006ChJAA...6..551Z};0
\citealp{2008ChJAA...8..411X};
\citealp{2010A&A...523A...5X}).

We estimate the X-ray ($0.3$-$10$keV) emission of the external-forward shock
for our obtained structured jet in Figure \ref{myFigC}.
The obtained X-ray light curves are shown in Figure \ref{myFigC} with solid lines,
where the same total energy of the structured jet is adopted.
That is to say, the values of $\varepsilon_{\max}=4.17\times10^{53}\rm erg$, $1.91\times10^{53}\rm erg$, and $7.0\times 10^{52}\rm erg$ are adopted for the situations
with $\theta_{\rm pre}=2.5^{\circ}$, $5^{\circ}$, and $10^{\circ}$, respectively.
The dynamics of external-forward shock can refer to
\cite{1999MNRAS.309..513H},
and the value of other parameters to calculate the afterglow emission
are the electron equipartition parameter $\epsilon_{e}=0.1$,
the magnetic equipartition parameter $\epsilon_{B}=0.01$,
the electron power-law index $p=2.5$,
the interstellar medium density $n=0.01$,
the Lorentz factor $\Gamma(\theta, \phi)=200$,
and the luminosity distance $D_{\rm L}=40$~Mpc.
The afterglows are obtained by summing the resulting emission
of $500\times200$ grids (see Section~2) with different inclination to the light-of-sight.
The viewing angle $(\theta_{\rm {obs}},\phi_{\rm {obs}})$
with $\theta_{\rm {obs}}=0^{\rm \circ}$, $5^{\rm \circ}$, $10^{\rm \circ}$,
$20^{\rm \circ}$, $40^{\rm \circ}$, and $60^{\rm \circ}$ are shown with
red, green, blue, cyan, magenta, and yellow lines, respectively.
Here, the value of $\phi_{\rm {obs}}$ does not affect the profile of light curves
and $\phi_{\rm {obs}}=0^\circ$ is adopted.
For comparison,
the X-ray afterglows from the situation with $\theta_{\rm pre}=0$ is also
shown with ``$\circ$'' in Figure \ref{myFigC}
and the afterglows with the same viewing angle are plotted with the same color.
According to Figure \ref{myFigC}, the light curve of afterglows can be very different for different $\theta_{\rm pre}$, especially for the situation with $\theta_{\rm pre}\gtrsim \theta_{\rm open}$.
In addition, the flux in the normal decay phase of afterglows decreases
with increasing the value of $\theta_{\rm pre}$
for situations with $\theta_{\rm pre}<\theta_{\rm open}$ and $\theta_{\rm {obs}}=0^{\rm \circ}$.
This behavior may affect the estimated kinetic energy of the external shock.

In the following part, we study the jet structure for a precessing jet with an evolving $P(t)$.
The obtained jet structure would depend on the precession period $\tau$
and the evolution behavior of the jet power $P(t)$.
For a jet powered via \cite{Blandford_RD-1977-Znajek_RL_-MNRAS.179.433B} mechanism,
the jet power depends on the black hole (BH) mass $M$, the BH spin $a$, and the magnetic field $B$
accumulated near the BH horizon.
Since the magnetic field on the BH is supported by the accretion disk $\dot{M}$,
it is reasonable to assume $B^2\propto\dot{M}$ (\citealp{2013ApJ...767L..36W}).
If $M$ and $a$ remain constant,
the evolution of jet power can be described as
(e.g., \citealp{2013ApJ...767L..36W}; \citealp{2015MNRAS.446.3642Y})
\begin{eqnarray}\label{Eq:LC_fit}
P_{\rm jet}(t)\sim P_{\rm 0}\left[\frac{1}{2}\left(\frac{t}{t_{\rm r}}\right)^{-a_{\rm r}s}+
\frac{1}{2}\left(\frac{t}{t_{\rm d}}\right)^{-a_{\rm d}s}\right]^{-1/s}
\end{eqnarray}
with $P_{\rm 0}=10^{52}\rm erg/s$, $a_{\rm r}=1/2$, $a_{\rm d}=-5/3$, $s=6$, $t_{\rm r}=0.1\tau$,
and $t_{\rm end}=10\tau$.
We adopt the following two case to discuss the dependence of the jet structure on
$\tau$ and $P(t)$: (I) $t_{\rm d}=0.1\tau$, (II) $t_{\rm d}=\tau$.
The obtained energy distribution is shown in Figure \ref{myFigD},
where $\theta_{\rm pre}=2.5^{\rm \circ}$, $5^{\rm \circ}$, and $10^{\rm \circ}$
are adopted in the left, middle, and right panels, respectively.
One can find that the jet structure is very complex in these situations.
Moreover, the structured jet becomes uniform in $\phi$-direction if
the evolution timescale of $P(t)$ is at around or larger than the precession period $\tau$.
The X-ray afterglows are also calculated and shown in Figure \ref{myFigE},
where $\theta_{\rm {obs}}=0^{\rm \circ}$ (red lines), $5^{\rm \circ}$ (green lines), $10^{\rm \circ}$ (blue lines),
$20^{\rm \circ}$ (cyan lines), $40^{\rm \circ}$ (magenta lines), and $60^{\rm \circ}$ (yellow lines)
with $\phi_{\rm {obs}}=0^{\rm \circ}$ (solid lines) or $180^{\rm \circ}$ (dashed lines) are adopted.
The situation with $\theta_{\rm {obs}}=0^{\rm \circ}$ and $\phi_{\rm {obs}}=0^{\rm \circ}$ is the same as that
with $\theta_{\rm {obs}}=0^{\rm \circ}$ and $\phi_{\rm {obs}}=180^{\rm \circ}$.
Then, we only plot the solid line for the situation with $\theta_{\rm {obs}}=0^{\rm \circ}$.
Here, the ''$\circ$'' is the same as Figure \ref{myFigC}.
One can find that the afterglows with $\theta_{\rm pre}\neq 0^\circ$ and those with $\theta_{\rm pre}=0^\circ$
are very different.
Moreover, the observer with different azimuthal angle $\phi_{\rm obs}$
may detect different light curves of afterglows.

\section{Conclusions and Discussions}\label{Sec:Conclusion_and_Discussion}

\begin{figure*}\centering
\centering
\begin{tabular}{cc}
\includegraphics[width=.45\textwidth]{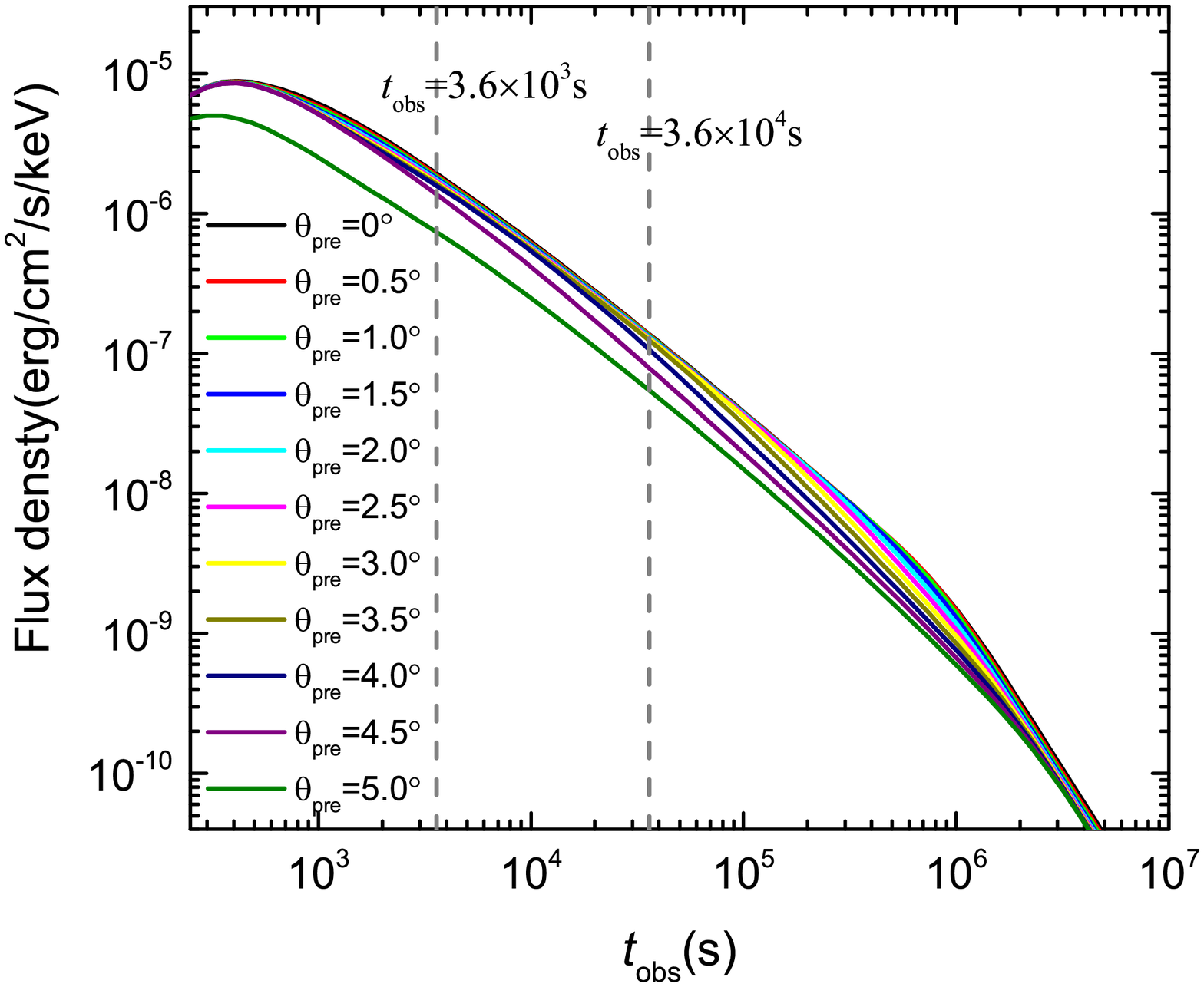}&
\includegraphics[width=.45\textwidth]{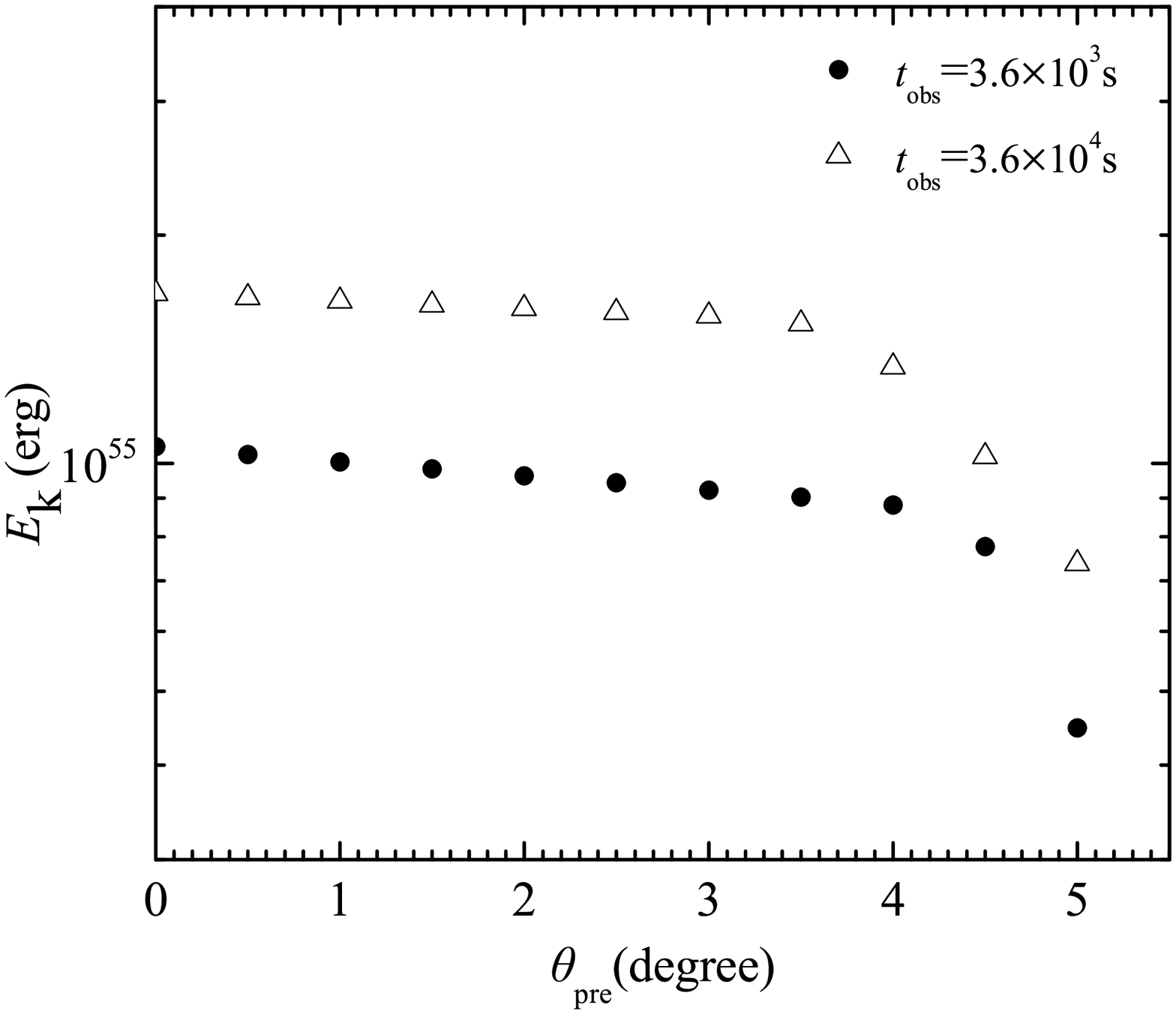}\\
\end{tabular}
\caption{The estimated kinetic energy (right panel) of jets based on X-ray afterglows (left panel) at observed time $t_{\rm obs
}$ for the situations with different $\theta_{\rm pre}$.}
\label{myFigF}
\end{figure*}
Most structured jet models used to explain the GRB~170817 afterglow can not and should not assume
that the structure is the same as that when the GRB is emitted.
In this work, we point out that
the jet structure in the prompt emission phase can be very different from that in the afterglow phase
for GRBs with a precessing jet.
For GRBs with a narrow-uniform-precessing jet,
a structured jet is difficult to form in the prompt emission phase
due to the low frequency of mergers between jet shells.
However, the structured jet can be easily formed in the early phase of afterglow.
We estimate the jet structure in the afterglow phase under the situation
that a narrow-uniform-precessing jet is launched from the central engine of a GRB.
With different precession angle,
the obtained structured jet can be roughly described as
a narrow uniform core with power-law wings and sharp cut-off edges,
a Gaussian profile, or a ring shape.
For a precessing jet with an evolving jet power $P(t)$, the obtained jet structure is very complex
and depends on both the precession period $\tau$ and the evolution behavior of $P(t)$.
In addition, the structured jet may be not axisymmetric.
The structured jet formed due to the precession of jet
is likely to be revealed by future observations for
a fraction of GW detected merging compact binary systems, e.g., BH-NS mergers (\citealp{Stone_N-2013-Loeb_A-PhRvD.87h4053S}).

We also calculate the X-ray emission of the external-forward shock for our obtained structured jet.
Our results show that the X-ray flux decreases with increasing the precession angle.
This can be found in the left and middle sub-figures of Figure \ref{myFigC} with $\theta_{\rm {obs}}\sim 0^\circ$.
We estimate the kinetic energy of the jet
based on the X-ray afterglows for situations with different $\theta_{\rm pre}$ ($<\theta_{\rm open}$)
and $\theta_{\rm {obs}}=0^\circ$.
The result is shown in Figure~\ref{myFigF},
where the left panel shows our synthetic light curves of afterglow emission at $10$~keV and the right panel plots the relation of $E_{\rm k}$ and $\theta_{\rm pre}$.
From this figure, one can find that the kinetic energy decreases with $\theta_{\rm pre}$.
Then, the larger value of $\theta_{\rm pre}$ is, the higher radiation efficiency of the GRB prompt emission would be found.
This may explain the exorbitant higher radiation efficiency of prompt emission found in some GRBs
(e.g., \citealp{2006MNRAS.370.1946G};
\citealp{2006A&A...458....7I};
\citealp{2006ApJ...642..389N};
\citealp{2007ApJ...655..989Z}).
The precession of a jet would also lead to incorrect estimates of the electron index.
In the left panel of Figure~\ref{myFigF},
the slope of the afterglow flux in the normal decay phase becomes steep
with the increase of the precession angle.
Then, the electron index estimated based on the slope of the normal decay phase
(\citealp{Zhang_B-2006-Fan_YZ-ApJ.642.354Z})
would become higher with the increase of the precession angle,
even though the intrinsic value of the electron index remains constant.
In other words, the spectral index and the decay slope of the normal decay phase
may deviate from the closure relations of the external-forward shock model (e.g., \citealp{Zhang_B-2004-Meszaros_P-IJMPA.19.2385Z,Zhang_B-2006-Fan_YZ-ApJ.642.354Z})
for GRBs with higher precession angle.
In addition, the time of jet-break caused by the edge effect
is gradually deferred with the increase of the precession angle.
The jet break even becomes unclear in the situation with high precession angle.
This behavior may help to understand the lack of expected jet breaks in \emph{Swift} X-ray afterglows
(e.g., \citealp{Racusin_JL-2009-Liang_EW-ApJ.698.43R, Wang_XiangGao-2018-Zhang_Bing-ApJ.859.160W}).
The incorrect estimates about the electron index and the deferred jet break time
may affect the estimated GRB total output energy.

\section*{Acknowledgments}
We thank the anonymous referee of this work for beneficial
suggestions that improved the paper.
This work is supported by the National Natural Science Foundation of China
(grant Nos. 11773007, 11533003, 11673006, 11822304, U1731239), the Guangxi Science Foundation (grant Nos. 2018GXNSFFA281010, 2016GXNSFDA380027, 2017AD22006, 2016GXNSFFA380006, 2018GXNSFGA281005), the Innovation Team and Outstanding Scholar Program in Guangxi Colleges, and the One-Hundred-Talents Program of Guangxi colleges.


\bsp	
\label{lastpage}

\begin{thebibliography}{99}
\bibitem[\protect\citeauthoryear{Abbott et al.}{2017a}]{2017ApJ...850L..39A} Abbott B.~P., et al., 2017a, ApJ, 850, L39
\bibitem[\protect\citeauthoryear{Abbott et al.}{2017b}]{2017ApJ...850L..40A} Abbott B.~P., et al., 2017b, ApJ, 850, L40
\bibitem[\protect\citeauthoryear{Abbott et al.}{2017c}]{2017ApJ...841...89A} Abbott B.~P., et al., 2017c, ApJ, 841, 89
\bibitem[\protect\citeauthoryear{Abbott et al.}{2017d}]{2017PhRvD..96b2001A} Abbott B.~P., et al., 2017d, PhRvD, 96, 022001
\bibitem[\protect\citeauthoryear{Arcavi et al.}{2017}]{2017Natur.551...64A} Arcavi I., et al., 2017, Natur, 551, 64
\bibitem[\protect\citeauthoryear{Berger, Fong, \& Chornock}{2013}]{2013ApJ...774L..23B} Berger E., Fong W., Chornock R., 2013, ApJ, 774, L23
\bibitem[\protect\citeauthoryear{Blackman, Yi, \& Field}{1996}]{1996ApJ...473L..79B} Blackman E.~G., Yi I., Field G.~B., 1996, ApJ, 473, L79
\bibitem[\protect\citeauthoryear{Blandford \& Znajek}{1977}]{Blandford_RD-1977-Znajek_RL_-MNRAS.179.433B} Blandford R.~D., Znajek R.~L., 1977, MNRAS, 179, 433
\bibitem[\protect\citeauthoryear{Coulter et al.}{2017}]{2017Sci...358.1556C} Coulter D.~A., et al., 2017, Sci, 358, 1556
\bibitem[\protect\citeauthoryear{Dai \& Gou}{2001}]{2001ApJ...552...72D} Dai Z.~G., Gou L.~J., 2001, ApJ, 552, 72
\bibitem[\protect\citeauthoryear{D'Avanzo et al.}{2018}]{2018A&A...613L...1D} D'Avanzo P., et al., 2018, A\&A, 613, L1
\bibitem[\protect\citeauthoryear{Evans et al.}{2017}]{2017Sci...358.1565E} Evans P.~A., et al., 2017, Sci, 358, 1565
\bibitem[\protect\citeauthoryear{Fern{\'a}ndez \& Metzger}{2016}]{2016ARNPS..66...23F} Fern{\'a}ndez R., Metzger B.~D., 2016, ARNPS, 66, 23
\bibitem[\protect\citeauthoryear{Foucart et al.}{2011}]{2011PhRvD..83b4005F} Foucart F., Duez M.~D., Kidder L.~E., Teukolsky S.~A., 2011, PhRvD, 83, 024005
\bibitem[\protect\citeauthoryear{Fraija et al.}{2017}]{2017arXiv171008514F} Fraija N., De Colle F., Veres P., Dichiara S., Barniol Duran R., Galvan-Gamez A., 2017, arXiv, arXiv:1710.08514
\bibitem[\protect\citeauthoryear{Goldstein et al.}{2017}]{2017ApJ...848L..14G} Goldstein A., et al., 2017, ApJ, 848, L14
\bibitem[\protect\citeauthoryear{Gottlieb et al.}{2018}]{2018MNRAS.479..588G} Gottlieb O., Nakar E., Piran T., Hotokezaka K., 2018, MNRAS, 479, 588
\bibitem[\protect\citeauthoryear{Granot, K{\"o}nigl, \& Piran}{2006}]{2006MNRAS.370.1946G} Granot J., K{\"o}nigl A., Piran T., 2006, MNRAS, 370, 1946
\bibitem[\protect\citeauthoryear{Granot}{2005}]{2005ApJ...631.1022G} Granot J., 2005, ApJ, 631, 1022
\bibitem[\protect\citeauthoryear{Haggard et al.}{2017}]{2017ApJ...848L..25H} Haggard D., Nynka M., Ruan J.~J., Kalogera V., Cenko S.~B., Evans P., Kennea J.~A., 2017, ApJ, 848, L25
\bibitem[\protect\citeauthoryear{Hallinan et al.}{2017}]{2017Sci...358.1579H} Hallinan G., et al., 2017, Sci, 358, 1579
\bibitem[\protect\citeauthoryear{Hinderer et al.}{2018}]{2018arXiv180803836H} Hinderer T., et al., 2018, arXiv, arXiv:1808.03836
\bibitem[\protect\citeauthoryear{Hu et al.}{2017}]{2017SciBu..62.1433H} Hu L., et al., 2017, SciBu, 62, 1433
\bibitem[\protect\citeauthoryear{Huang, Dai, \& Lu}{1999}]{1999MNRAS.309..513H} Huang Y.~F., Dai Z.~G., Lu T., 1999, MNRAS, 309, 513
\bibitem[\protect\citeauthoryear{Ioka et al.}{2006}]{2006A&A...458....7I} Ioka K., Toma K., Yamazaki R., Nakamura T., 2006, A\&A, 458, 7
\bibitem[\protect\citeauthoryear{Kasen et al.}{2017}]{2017Natur.551...80K} Kasen D., Metzger B., Barnes J., Quataert E., Ramirez-Ruiz E., 2017, Natur, 551, 80
\bibitem[\protect\citeauthoryear{Kasliwal et al.}{2017}]{2017Sci...358.1559K} Kasliwal M.~M., et al., 2017, Sci, 358, 1559
\bibitem[\protect\citeauthoryear{Kisaka et al.}{2018}]{2018ApJ...867...39K} Kisaka S., Ioka K., Kashiyama K., Nakamura T., 2018, ApJ, 867, 39
\bibitem[\protect\citeauthoryear{Kumar \& Granot}{2003}]{2003ApJ...591.1075K} Kumar P., Granot J., 2003, ApJ, 591, 1075
\bibitem[\protect\citeauthoryear{Lamb \& Kobayashi}{2017}]{2017MNRAS.472.4953L} Lamb G.~P., Kobayashi S., 2017, MNRAS, 472, 4953
\bibitem[\protect\citeauthoryear{Lamb \& Kobayashi}{2018}]{2018MNRAS.478..733L} Lamb G.~P., Kobayashi S., 2018, MNRAS, 478, 733
\bibitem[\protect\citeauthoryear{Lamb, et al.}{2019}]{Lamb_GP-2019-Lyman_JD-ApJ.870L.15L} Lamb G.~P., et al., 2019, ApJ, 870, L15
\bibitem[\protect\citeauthoryear{Lazzati et al.}{2018}]{2018PhRvL.120x1103L} Lazzati D., Perna R., Morsony B.~J., Lopez-Camara D., Cantiello M., Ciolfi R., Giacomazzo B., Workman J.~C., 2018, PhRvL, 120, 241103
\bibitem[\protect\citeauthoryear{Lei et al.}{2007}]{2007A&A...468..563L} Lei W.~H., Wang D.~X., Gong B.~P., Huang C.~Y., 2007, A\&A, 468, 563
\bibitem[\protect\citeauthoryear{Li \& Paczy{\'n}ski}{1998}]{1998ApJ...507L..59L} Li L.-X., Paczy{\'n}ski B., 1998, ApJ, 507, L59
\bibitem[\protect\citeauthoryear{Lin et al.}{2018}]{2018ApJ...856...90L} Lin D.-B., Liu T., Lin J., Wang X.-G., Gu W.-M., Liang E.-W., 2018, ApJ, 856, 90
\bibitem[\protect\citeauthoryear{Lipunov et al.}{2017}]{2017ApJ...850L...1L} Lipunov V.~M., et al., 2017, ApJ, 850, L1
\bibitem[\protect\citeauthoryear{Liska et al.}{2018}]{2018MNRAS.474L..81L} Liska M., Hesp C., Tchekhovskoy A., Ingram A., van der Klis M., Markoff S., 2018, MNRAS, 474, L81
\bibitem[\protect\citeauthoryear{Liu, Gu, \& Zhang}{2017}]{2017NewAR..79....1L} Liu T., Gu W.-M., Zhang B., 2017, NewAR, 79, 1
\bibitem[\protect\citeauthoryear{Lyman et al.}{2018}]{Lyman_JD-2018-Lamb_GP-NatAs.2.751L} Lyman J.~D., et al., 2018, NatAs, 2, 751
\bibitem[\protect\citeauthoryear{Margutti et al.}{2018}]{2018ApJ...856L..18M} Margutti R., et al., 2018, ApJ, 856, L18
\bibitem[\protect\citeauthoryear{Meng et al.}{2018}]{2018ApJ...860...72M} Meng Y.-Z., et al., 2018, ApJ, 860, 72
\bibitem[\protect\citeauthoryear{Metzger \& Berger}{2012}]{2012ApJ...746...48M} Metzger B.~D., Berger E., 2012, ApJ, 746, 48
\bibitem[\protect\citeauthoryear{Mooley et al.}{2018}]{2018Natur.554..207M} Mooley K.~P., et al., 2018, Natur, 554, 207
\bibitem[\protect\citeauthoryear{Murguia-Berthier et al.}{2017}]{2017ApJ...848L..34M} Murguia-Berthier A., et al., 2017, ApJ, 848, L34
\bibitem[\protect\citeauthoryear{Nousek et al.}{2006}]{2006ApJ...642..389N} Nousek J.~A., et al., 2006, ApJ, 642, 389
\bibitem[\protect\citeauthoryear{Pian et al.}{2017}]{2017Natur.551...67P} Pian E., et al., 2017, Natur, 551, 67
\bibitem[\protect\citeauthoryear{Portegies Zwart, Lee, \& Lee}{1999}]{1999ApJ...520..666P} Portegies Zwart S.~F., Lee C.-H., Lee H.~K., 1999, ApJ, 520, 666
\bibitem[\protect\citeauthoryear{Racusin, et al.}{2009}]{Racusin_JL-2009-Liang_EW-ApJ.698.43R} Racusin J.~L., et al., 2009, ApJ, 698, 43
\bibitem[\protect\citeauthoryear{Resmi et al.}{2018}]{2018ApJ...867...57R} Resmi L., et al., 2018, ApJ, 867, 57
\bibitem[\protect\citeauthoryear{Reynoso, Romero, \& Sampayo}{2006}]{2006A&A...454...11R} Reynoso M.~M., Romero G.~E., Sampayo O.~A., 2006, A\&A, 454, 11
\bibitem[\protect\citeauthoryear{Rossi, Lazzati, \& Rees}{2002}]{2002MNRAS.332..945R} Rossi E., Lazzati D., Rees M.~J., 2002, MNRAS, 332, 945
\bibitem[\protect\citeauthoryear{Ruan et al.}{2018}]{2018ApJ...853L...4R} Ruan J.~J., Nynka M., Haggard D., Kalogera V., Evans P., 2018, ApJ, 853, L4
\bibitem[\protect\citeauthoryear{Savchenko et al.}{2016}]{2016ApJ...820L..36S} Savchenko V., et al., 2016, ApJ, 820, L36
\bibitem[\protect\citeauthoryear{Smartt et al.}{2017}]{2017Natur.551...75S} Smartt S.~J., et al., 2017, Natur, 551, 75
\bibitem[\protect\citeauthoryear{Soares-Santos et al.}{2017}]{2017ApJ...848L..16S} Soares-Santos M., et al., 2017, ApJ, 848, L16
\bibitem[\protect\citeauthoryear{Song, Liu, \& Li}{2018}]{2018MNRAS.477.2173S} Song C.-Y., Liu T., Li A., 2018, MNRAS, 477, 21730
\bibitem[\protect\citeauthoryear{Stone, Loeb, \& Berger}{2013}]{Stone_N-2013-Loeb_A-PhRvD.87h4053S} Stone N., Loeb A., Berger E., 2013, PhRvD, 87, 084053
\bibitem[\protect\citeauthoryear{Tanvir et al.}{2017}]{2017ApJ...848L..27T} Tanvir N.~R., et al., 2017, ApJ, 848, L27
\bibitem[\protect\citeauthoryear{Troja et al.}{2017}]{2017Natur.551...71T} Troja E., et al., 2017, Natur, 551, 71
\bibitem[\protect\citeauthoryear{Troja et al.}{2018}]{2018MNRAS.478L..18T} Troja E., et al., 2018, MNRAS, 478, L18
\bibitem[\protect\citeauthoryear{Valenti et al.}{2017}]{2017ApJ...848L..24V} Valenti S., et al., 2017, ApJ, 848, L24
\bibitem[\protect\citeauthoryear{van Eerten, et al.}{2018}]{vanEerten_ETH-2018-Ryan_G-arXiv180806617V} van Eerten E.~T.~H., et al., 2018, arXiv e-prints, arXiv:1808.06617
\bibitem[\protect\citeauthoryear{Wang, et al.}{2018}]{Wang_XiangGao-2018-Zhang_Bing-ApJ.859.160W} Wang X.-G., Zhang B., Liang E.-W., Lu R.-J., Lin D.-B., Li J., Li L., 2018, ApJ, 859, 160
\bibitem[\protect\citeauthoryear{Wu, Hou, \& Lei}{2013}]{2013ApJ...767L..36W} Wu X.-F., Hou S.-J., Lei W.-H., 2013, ApJ, 767, L36
\bibitem[\protect\citeauthoryear{Xiao et al.}{2017}]{2017ApJ...850L..41X} Xiao D., Liu L.-D., Dai Z.-G., Wu X.-F., 2017, ApJ, 850, L41
\bibitem[\protect\citeauthoryear{Xu \& Huang}{2010}]{2010A&A...523A...5X} Xu M., Huang Y.~F., 2010, A\&A, 523, A5
\bibitem[\protect\citeauthoryear{Xu, Huang, \& Kong}{2008}]{2008ChJAA...8..411X} Xu M., Huang Y.-F., Kong S.-W., 2008, ChJAA, 8, 411
\bibitem[\protect\citeauthoryear{Yu et al.}{2015}]{2015MNRAS.446.3642Y} Yu Y.~B., Wu X.~F., Huang Y.~F., Coward D.~M., Stratta G., Gendre B., Howell E.~J., 2015, MNRAS, 446, 3642
\bibitem[\protect\citeauthoryear{Zhang \& M{\'e}sz{\'a}ros}{2002}]{2002ApJ...571..876Z} Zhang B., M{\'e}sz{\'a}ros P., 2002, ApJ, 571, 876
\bibitem[\protect\citeauthoryear{Zhang \& M{\'e}sz{\'a}ros}{2004}]{Zhang_B-2004-Meszaros_P-IJMPA.19.2385Z} Zhang B., M{\'e}sz{\'a}ros P., 2004, IJMPA, 19, 2385
\bibitem[\protect\citeauthoryear{Zhang \& Yan}{2011}]{2011ApJ...726...90Z} Zhang B., Yan H., 2011, ApJ, 726, 90
\bibitem[\protect\citeauthoryear{Zhang et al.}{2006}]{Zhang_B-2006-Fan_YZ-ApJ.642.354Z} Zhang B., Fan Y.~Z., Dyks J., Kobayashi S., M{\'e}sz{\'a}ros P., Burrows D.~N., Nousek J.~A., Gehrels N., 2006, ApJ, 642, 354
\bibitem[\protect\citeauthoryear{Zhang et al.}{2007}]{2007ApJ...655..989Z} Zhang B., et al., 2007, ApJ, 655, 989
\bibitem[\protect\citeauthoryear{Zhang, et al.}{2018}]{Zhang_BB-2018-Zhang_B-NatCo.9.447Z} Zhang B.-B., et al., 2018, NatCo, 9, 447
\bibitem[\protect\citeauthoryear{Zou \& Dai}{2006}]{2006ChJAA...6..551Z} Zou Y.-C., Dai Z.-G., 2006, ChJAA, 6, 551
\end{thebibliography}
\end{document}